\documentclass[fleqn,nofootinbib,showkeys,pra]{elsarticle}
\usepackage{setspace}
\newcommand{\R}{{\mathbb{R}}}

\newcommand{\I}{{\mathbb{I}}}
\usepackage{amssymb}
\usepackage{xcolor}

\begin{document}
\title{Conditions for realizing one-point interactions from a 
 multi-layer structure model
}
\author{Alexander V. Zolotaryuk and Yaroslav Zolotaryuk}
\address
{Bogolyubov Institute for Theoretical Physics, National Academy of
Sciences of Ukraine, Kyiv 03143, Ukraine}


\date{\today}

\begin{abstract}

A  heterostructure composed of $N$ parallel homogeneous layers is studied 
in the  limit as their widths $l_1, \ldots , l_N$  shrink to zero.
 The problem is investigated in one dimension and
the piecewise constant potential in the Schr\"{o}dinger equation is given by
the  strengths  $V_1, \ldots , V_N$ as functions of $l_1, \ldots , l_N$,
respectively. The key point is the derivation of the conditions on the
  functions $V_1(l_1), \ldots , V_N(l_N)$  for realizing a family of one-point 
  interactions as $l_1, \ldots , l_N$  tend to zero along available paths
in the $N$-dimensional space.  
       The existence of equations for a squeezed structure, the solution of which 
determines the system parameter values, under    which the non-zero tunneling 
of quantum particles through a multi-layer structure occurs, is shown to  exist 
and depend on the paths. This tunneling 
  appears as a result of an appropriate cancellation of divergences.

  \end{abstract}

\bigskip


\maketitle


\section{Introduction}

The differential operators with singular zero-range potentials
   are frequently used to model different physical systems.
The particular class of such models is 
represented by Schr\"{o}dinger operators in one dimension with regular realistic potentials,
which can be approximated by point (called also contact) interactions 
 (see books \cite{a-h,ak} for details and references)
defined on the sets consisting of isolated points, curves or surfaces.
In this asymptotic approach, for a general class of potentials $V$,
under certain conditions the functions $\varepsilon^{-1}V(x/\varepsilon)$
and $\varepsilon^{-2}V(x/\varepsilon)$ converge in the sense of distributions
as $\varepsilon \to 0$  to the Dirac delta-function $\delta(x)$ and its first derivative
 $\delta'(x)$, respectively, realizing one-point interactions.
Thus, the one-dimensional Schr\"{o}dinger operator
with the $\delta'(x)$ distribution has rigorously been treated in works
\cite{s,gm,gh,gh1} and afterwards these studies were extended to the quantum graph
theory \cite{m1,m2,em1}.
Some particular cases of the regularization of the $\delta'(x)$ function
for realizing one-point interactions have been used in \cite{tn}.
Using a special class of piecewise constant potentials $V$, 
it was possible \cite{c-g,zci,asl} to construct explicitly the matrices,  
which connect the two-sided boundary conditions at the point of singularity.

In a similar way, the Schr\"{o}dinger operator with the combined potential 
$a \delta'(x) + b \delta(x)$ was investigated \cite{zz11jpa,g1,g2}, 
including the studies of the discrete spectrum \cite{gnn,gmmn}.
The one-dimensional quantum Hamiltonians with potentials
having a regular unperturbed part of different shape 
 and a singular perturbation of the form
$a \delta'(x) + b \delta(x)$ have been examined in publications \cite{ggn,gggm,fggn1}.

Various aspects have been studied in numerous works (see, e.g., 
publications \cite{s,adk,an1,l1,bn,an2,l2,g3},
a few to mention) concerning the $\delta'$-interaction, which
should not be confused with the  $\delta'$-potential. It is worth mentioning the works
on the three delta approximation of the $\delta'$-interaction  \cite{cs,enz}
 as well as publications \cite{fggn1,afr1,afr2}, where the $\delta'$-interaction was 
 considered as a perturbation of regular potentials. 

The most simple example of regularizing the $\delta'$-potential is the sequence 
of piecewise constant functions, which from the physical point of view
 can be related to planar heterostructures.  In particular, 
 for the  double-layer case, using 
 a three-scale parametrization of the layer parameters such as their strengths, 
 thickness and the distance between the layers, a whole family of one-point
interactions has been constructed, where the scattering data \cite{asl,z18pe,z18aop}
  as well as the existence of a bound state \cite{zz20ltp,zz21jpa} 
  have been  examined. In general, here 
  it is not required that a piecewise constant potential 
 must have a well-defined limit (for instance,  in the sense of distributions) 
 as the thickness of layers and the distance between them  shrink to zero.
As a result, within the  three-scaled squeezing procedure, it was possible to    
describe both the {\it resonant}  point interactions \cite{gm,gh,gh1,tn,c-g,zz11jpa,g1,g2}  
and the point interaction produced by the potential $a \delta'(x)$ potential, 
that comes from the theory developed by Kurasov and coworkers \cite{a-h,k,cnf}, for which
 the elements of the diagonal connection matrix are $\theta = (2 +a)/(2-a)$ and 
 $\theta^{-1}$  \cite{gnn}, within an unique scheme \cite{z18aop}.

The present paper is devoted to the investigation of a planar 
heterostructure composed of an arbitrary number of
  layers. The layers may be   separated by  distances
 and these spaces (if any) are considered as a particular case of layers
 with zero strength potential. The goal of the paper is to study this 
 multi-layer structure
  in the limit as both the layer thickness and the distance between the layers  
shrink to zero.  The electron motion in this system is supposed to be 
 confined in the longitudinal
direction (say, along the $x$-axis) being perpendicular to the transverse planes 
where electronic motion is free. Then 
the three-dimensional Schr\"{o}dinger equation of such a structure can be separated
into longitudinal and transverse parts, resulting in  the reduced 
stationary one-dimensional Schr\"{o}dinger equation 
\begin{equation}
-\, \psi''(x) +V(x)\psi(x)= E\psi(x)
\label{1}
\end{equation}
with respect to the longitudinal component of the
wave function $\psi(x)$ and the electron energy $E$. Here, $V(x)$ is 
the potential for quantum particles in the form of a piecewise constant function 
 and the prime stands for the differentiation over $x$.

\section{Transmission matrix for a multilayer structure}

Consider   one-dimensional stationary Schr\"{o}dinger equation (\ref{1}),
where the potential $V(x)$ is a piecewise constant function defined on 
the interval $(x_0, x_N)$ with $N$ subsets 
$(x_{i-1}, x_{i})$, where $i= 1, \ldots , N$ and $N = 1, 2, \ldots$ .
Each strength $ V_i \in \R$ defined on this interval is supposed to depend 
on the $i$th layer width $l_i := x_i - x_{i-1}$, so that we have
the set of functions: $V_1(l_1), \ldots , V_{N}(l_N)$.

The transmission matrix $\Lambda_i(x_{i-1}, x_i)$ for the $i$th layer
connects the values of the wave function $\psi_i(x)$
and its derivative $\psi'_i(x)$ at the boundaries $x =x_{i-1}$ and $x=x_i$
according to the matrix equation
\begin{equation}
\!\!\!\!\!\!\!\!\!\!\!\!\!
 \left(\begin{array}{cc}\!
\psi_i(x_i) \\ \psi'_i(x_i) \end{array}\! \right)= \Lambda_i(x_{i-1}, x_i)
\left(\! \begin{array}{cc}
\psi_i(x_{i-1}) \\ \psi'_i(x_{i-1}) \end{array} \!\right),~
\Lambda_i(x_{i-1}, x_i) = \left( \!\begin{array}{cc} \lambda_{i,11}~~\lambda_{i,12}\\
\lambda_{i,21}~~\lambda_{i,22} \end{array}\! \right).
\label{2}
\end{equation}
The transmission matrix for each layer defined on the interval 
$(x_{i-1}, x_i)$ can be computed through the solutions
$u_i(x)$ and $v_i(x)$ and their derivatives taken at the boundaries $x=x_{i-1}$ 
and $x=x_i$. Let $u_i(x)$ and $v_i(x)$ be linearly independent solutions on the interval 
$(x_{i-1}, x_i)$ obeying the initial conditions 
\begin{equation}
u_i(x_{i-1})=1,~~u'_i(x_{i-1})=0,~~v_i(x_{i-1})=0,~~v'_i(x_{i-1})=1.
\label{3}
\end{equation}
 Then the representation of the $\Lambda_i$-matrix reads
\begin{equation}
\Lambda_i(x_{i-1}, x_i) = \left( \begin{array}{cc} u_i(x_i) ~~v_i(x_i) \\
 u'_i(x_i) ~~v'_i(x_i) \end{array} \right), ~~~\det\Lambda_i(x_{i-1}, x_i)=1.
 \label{4} 
\end{equation}
Particularly, for the piecewise constant function  $V_i(x)$, 
the solutions $u_i(x)$ and $v_i(x)$,
$x_{i-1} \le x \le x_i$, obeying conditions (\ref{3}), read
\begin{equation}
u_i(x) = \cos[q_i(x-x_{i-1})],~~v_i(x) =q_i^{-1} \sin[q_i(x-x_{i-1})], 
\label{5}
\end{equation}
 where 
\begin{equation}
 q_i= \sqrt{E - V_i}\,.
 \label{6}
\end{equation}
Therefore the $\Lambda_i$-matrix in this particular case is
\begin{equation}
\Lambda_i(x_{i-1}, x_i) = \left(\, \begin{array}{cc} \cos(q_il_i)  ~~& q_i^{-1}\sin(q_il_i) \\
 -q_i \sin(q_il_i) ~~& \cos(q_il_i) \end{array} \right), ~~
 i = \overline{1,N},
 \label{7} 
\end{equation}
and the full transmission matrix,  connecting the boundary 
conditions at $x= x_0$ and $x=x_N$, is the product
\begin{equation}
\Lambda(x_0, x_N) = \Lambda_N(x_{N-1}, x_N)\ldots \Lambda_1(x_0, x_1) = 
 \left( \begin{array}{cc} \lambda_{11}~~\lambda_{12} \\ 
\lambda_{21}~~\lambda_{22} \end{array} \right).
\label{8}
\end{equation} 
 Clearly, due to the form of matrices (\ref{7}),  $\det\Lambda(x_0,x_N) =1.$  
 
\section{Structure of the full transmission matrix $\Lambda(x_0,x_N)$ 
}

 Using  matrix representation (\ref{7}) in product (\ref{8}), by induction
one can derive some properties of the $\Lambda(x_0, x_N)$-matrix elements. 
First of all  note that each   element of this matrix
consists of $2^{N-1}$ terms, each of which is the
$N$-multiple product of sines and cosines involved in  matrices 
(\ref{7}). By lengthy and straightforward computation one can derive 
by induction the structure of  these terms: 
\begin{equation}
\left\{ \begin{array}{llll}  \lambda_{11} \\
\lambda_{12} \\ \lambda_{21} \\ \lambda_{22} \end{array} \right\}
= \prod_{n=1}^N    \cos(q_nl_n)\times \left\{ \begin{array}{llll}  ~~\, 1 +{\cal Q}_{11} \,,\\
~~ \sum_{i=1}^Nq_i^{-1}t_i  + {\cal Q}_{12} \,, \\
-   \sum_{i=1}^Nq_it_i - {\cal Q}_{21} \, ,\\
 ~~ \, 1 +{\cal Q}_{22} \,, \end{array} \right.
 \label{9}
\end{equation}
where $ t_i := \tan(q_il_i)$ and 
\begin{equation}
{\cal Q}_{11}= \sum_{n=1}^{[N/2]}(-1)^n \sum_{1 = i_1 < j_1 < i_2 < j_2 < \ldots < i_n< j_n=N}
 D_{i_1j_1} \ldots D_{i_nj_n} ,
 \label{10}
\end{equation}
\begin{equation}
\!\!\!\!\!\!\!\!\!\!\!\!\!\!
{\cal Q}_{12}= \sum_{n=1}^{[(N-1)/2]}(-1)^n \sum_{1 = i_1 < j_1 <i_2 <j_2  < \ldots < 
i_n< j_n < k_n =N}  D_{j_1i_1} \ldots D_{j_ni_n} q_{k_n}^{-1}t_{k_n} ,~~
 \label{11}
\end{equation}
\begin{equation}
\!\!\!\!\!\!\!\!\!\!\!\!\!\!
{\cal Q}_{21}= \sum_{n=1}^{[(N-1)/2]}(-1)^n \sum_{1 = i_1 < j_1 <i_2 <j_2  < \ldots < 
i_n< j_n < k_n =N}  D_{i_1j_1} \ldots D_{i_nj_n} q_{k_n}t_{k_n} ,~~
\label{12}
\end{equation}
\begin{equation}
{\cal Q}_{22}=  \sum_{n=1}^{[N/2]} (- 1)^n \sum_{1 = i_1 < j_1 < i_2 < j_2 < \ldots
 < i_n< j_n=N}  D_{j_1i_1} \ldots D_{j_ni_n} ,
\label{13}
\end{equation}
with 
\begin{equation}
 D_{ij}:=(q_i/q_j)t_it_j, ~~i,j = \overline{1,N}~(i \neq j).
\label{14} 
\end{equation}
Here, $[N/2]$ and  $[(N-1)/2]$ stand for the integer parts of $N/2$ and
$(N-1)/2$, respectively. 

It is remarkable that all the $Q$-series appear in the form  
of   pairwise terms (\ref{14}), called  `dyads' from now on.
The $n$th group of terms in (\ref{10}) or (\ref{13}) consists of 
 ${ N \choose \, 2n \, } = 
\frac{N !}{ (2n)! (N-2n)!}$ summands. Thus, for $n=1$ there are $N(N-1)/2$ 
dyads (\ref{14}), where $i =1, \ldots , N-1$ and $j = 2, \dots , N$ 
with the ordering relation $i < j$. Next, the terms in the group with $n=2$
are of the form $D_{i_1j_1}D_{i_2j_2}$ 
in $\lambda_{11}$ and $D_{j_1i_1}D_{j_2i_2}$ 
in $\lambda_{22}$, where $i_1 =1, \ldots , N-3$, $j_1 = 2, \ldots , N-2$,
$i_2 =3, \ldots , N-1$ and $j_2 = 4, \ldots , N$ with the ordering relation 
$i_1 < j_1 < i_2 < j_2$. In a similar way, one can present
 the terms in the group with  $n=3$ and so on.

Similarly, the $n$th group of terms in (\ref{11}) and (\ref{12}) consists of
 ${N \choose \, 2n+1 \,} = \frac{N !}{ (2n +1)! (N-2n-1)!}$ summands. 
  Thus, for $n=1$ we have the sum 
$N(N-1)(N-2)/6$ terms in real-valued functions $V_i(l_i)$ the form of `triads'
 $ D_{ji} q_{k}^{-1}t_k
 = (q_j/q_iq_k)t_it_jt_k $ in $\lambda_{12}$ and  $ D_{ij} q_{k}t_k = 
 (q_iq_k/q_j)t_it_jt_k $
in $\lambda_{21}$, where $i = 1, \ldots , N-2$, $j = 2, \ldots , N-1$ and
$k = 3, \ldots , N$ ($i< j< k$). In a similar way, one can present
 the terms in the group with  $n=2$ and so on.

\section{Conditions  for realizing  
one-point interactions with finite diagonal elements in the connection matrix}

A family of {\it one}-point interactions can be realized from 
equation (\ref{1}) with a piecewise constant potential $V(x)$ 
defined on the interval $x_0 \le x \le x_N$ if all the $N$ layer widths 
$l_1, \ldots , l_N $ shrink to zero, setting then 
   $x_0 \to -\, 0$ and $x_N \to +\, 0$.  Any finite configuration of 
the layer widths can be associated with a point (vector) $l := \mbox{col}(l_1, \ldots , l_N)$
in the   orthogonal angle of dimension $N$ with the vertex at the origin $l=0$, 
so that the squeezing of the 
 full layer thickness $x_N-x_0$ to zero can be described as a {\it path} approaching
the origin $l =0$. At the same time, for the realization of a 
non-trivial point interaction, at least one of the potential strengths
 $V_1, \ldots , V_N$, $i = 1, \ldots ,N$, must tend to infinity with some rate. 
Otherwise, the transmission through the squeezed structure will be trivial, i.e.,
the limit transmission (connection) matrix will be just the identity.
In general, under the squeezing procedure, the  behavior of the  functions
 $V_1(l_1), \ldots , V_N(l_N)$ versus the paths approaching the origin $l=0$
are to be analyzed below. 

While the elements $\lambda_{11}$ and $\lambda_{22}$ [see  equations (\ref{10}) and (\ref{13})]
are composed of only  dyads (\ref{14}),  the summands in 
the off-diagonal elements contain additional factors: $q_k^{-1}t_k$
in $\lambda_{12}$ and $q_kt_k$ in $\lambda_{21}$. 
Therefore, for the existence of finite squeezed limits 
of the elements $\lambda_{11}$ and $\lambda_{22}$,  it is sufficient to
estimate only the dyads $D_{ij}$ and $D_{ji}$  ($i < j$). 
Next, according to (\ref{11}) and (\ref{12}), these estimates are to be used
for the analysis of the elements $\lambda_{12}$ and $\lambda_{21}$.
Hence, the squeezing analysis of the full $\Lambda$-matrix is reduced to studying 
the conditions on all the pairs of functions $ V_i(l_i)$ and $V_j(l_j)$ versus the paths 
projected onto the $(N-1)N/2$  $(l_i,l_j)$-faces of the $N$-dimensional orthogonal angle. 

\subsection{A $\delta$-like squeezing limit}

  Consider first the case with the asymptotic 
behavior $V_i(l_i)l_i \to c \in \R$ for all the functions $V_i(l_i)$  as $l_i \to 0$. 
Here and in the following $c$ stands for an arbitrary constant.
Assume that  the functions $V_i(l_i)$ belong to the sets  
\begin{equation}
{\cal G}_{i}^0 := \left\{ V_i \in \R \, \Big|\,
\lim_{l_i \to 0} V_i(l_i)\,l_i  = \alpha_i \in \R \right\}\!,~~~i = \overline{1,N}.
\label{15}
\end{equation}
Note that the free space $(x_{i-1}, x_i)$ with  $V_i \equiv 0$, is also treated  as 
a `layer' with strength zero.
 
For any two variables  $\beta_1$ and $\beta_2$, we write  $\beta_1 \sim \beta_2$
if they are of the same order.  Let us now estimate series (\ref{10})--(\ref{13}) 
in the case as all the functions $V_i$ belong to ${\cal G}_{i}^0$ ($i=1, \dots , N$).  
In the  limit as $l \to 0$, we have $D_{ij} \sim \alpha_il_j \to 0$
and $D_{ji} \sim \alpha_j l_i \to 0$ as well as 
 $ D_{ji}q_k^{-1}t_k \sim \alpha_j l_i l_k  \to 0$ and 
$ D_{ij}q_k t_k \sim \alpha_i l_j \alpha_k  \to 0$. Moreover, the shrinking of
all the widths $l_i$'s to zero is available to be arranged in a repeated manner 
(not simultaneously). 
For instance, the shrinking may pass along some  edges of the orthogonal angle or 
 be found in its ($l_i,l_j$)-faces. 
 Hence, all the paths with $l \to 0$ constitute in this case   
a {\it pencil} that coincides with the $N$-dimensional orthogonal angle.
Thus,  as follows from equations
(\ref{10})--(\ref{13}), the $Q$-series in equations (\ref{9}) vanish, resulting in
 the limit transmission (connection) matrix of the form
\begin{equation}
\Lambda_0 : = \lim_{l \to 0} \Lambda(x_0,x_N) = \left( \begin{array}{lr} 1 ~~0 \\
\alpha ~~1 \end{array} \right),~~~~\alpha := \sum_{i=1}^N \alpha_i\,.
\label{16}
\end{equation}
In the case if $\alpha > 0$, we have the point interaction specified by a 
$\delta$-potential barrier, while for $\alpha <0$, we
  are dealing with an effective $\delta$-well potential.
Using the general equation for bound state levels $\kappa $ ($ k  
= {\rm i}\kappa$, $\kappa >0$), written in terms of 
the $\Lambda$-matrix elements as (see, e.g., \cite{zz21jpa,zz14ijmpb}) 
\begin{equation}
\lambda_{11}(\kappa)+ \lambda_{22}(\kappa) + \kappa\lambda_{12}(\kappa) 
+ \kappa^{-1}\lambda_{21}(\kappa) = 0,
\label{17}
\end{equation}
we obtain $\kappa = -\, \alpha/2$.  Thus, one can formulate 
 
 {\bf Conclusion 1}\,: {\it Let all the functions $V_i(l_i)$, $i =1, \ldots , N$, 
 belong to the  $ {\cal G}_{i}^0$-sets defined by condition (\ref{15})
  and the vector $l= \mbox{col}(l_1, \ldots , l_N)$ 
 tends to the origin $l =0$ along any path within the pencil formed by 
  the $N$-dimensional orthogonal angle with the vertex at the origin $l=0$. 
  Then, in the squeezing limit  $l \to 0$, the point interactions
 are realized as the family of $\delta$-potentials with the connection matrix given by 
 equation (\ref{16}).}

\subsection{A singular squeezing limit}

 The set of functions $V_i(l_i) \in 
{\cal G}_{i}^0$, used for materializing  $\delta$-potentials, can be extended 
to include the functions with a more singular behavior like $|V_i(l_i)| \,l_i \to \infty$
as $l_i \to 0$. In the case if  at least one of the functions $V_i(l_i)$, 
$i = 1, \ldots ,N$, possesses  such an infinite behavior, the shrinking of the 
full system to zero dimension will be
referred to as a {\it singular} squeezing limit from here on.  

The sufficient requirement for realizing a point interaction is
that the arguments $q_il_i$ of the trigonometric functions 
   in matrix expressions (\ref{7})  must be finite  in the limit as $l_i \to 0$
even if the product  $V_i(l_i) \,l_i $ diverges as $l_i \to 0$.  
In order to derive the conditions under which the $l_i \to 0$ limit 
is finite,  we define the sets of  real-valued 
functions $V_i(l_i)$, $i = 1, \ldots , N$,  by the conditions
\begin{equation}
{\cal G}_i  :=  \left\{ V_i(l_i) \in \R ~\big|~ 
 \lim_{l_i \to 0}|V_i(l_i)|^{1/2}l_i =   0 \right\},   
 \label{18}
\end{equation} 
with the corresponding complements 
\begin{equation}
 {\cal G}'_i  :=  \left\{ V_i(l_i) \in \R ~\big|~ \lim_{l_i \to 0}|V_i(l_i)|^{1/2}l_i 
 = c > 0 \right\}. 
\label{19}
\end{equation}
 Hence, the trigonometric functions in expressions (\ref{9})--(\ref{14}) make sense if 
each function  $V_i(l_i)$ belongs to the union 
$ \bar{\cal G}_i: =  {\cal G}_i  \cup {\cal G}'_i$\,. 

The condition  $V_i(l_i) \in \bar{\cal G}_i$, $i=1, \ldots , N$, 
is not sufficient for the existence of finite squeezed limits
of the diagonal elements $\lambda_{11}$ and $\lambda_{22}$ 
because  of the presence of the  factors
$q_i$ with  $|q_i| \to \infty$ in series (\ref{10}) and (\ref{13}).  However, 
 this divergence can be suppressed due to the presence of the `neighboring' factors
 $q_j^{-1}$ ($j \neq i$), leading to the ambiguity of type  $\infty /\infty$.
Contrary to the $\delta$-potential case, here the finite limits are possible 
only if all the thickness parameters  $l_i$, $i=1, \ldots ,N$,
 converge to the origin  simultaneously.   Thus, 
 for the existence of finite squeezed limits 
of the elements $\lambda_{11}$ and $\lambda_{22}$, it is sufficient to
estimate both the dyads $D_{ij}$ and $D_{ji}$  ($i < j$). 
Due to definition (\ref{14}),  the asymptotic expressions for these dyads,  
written in terms of the functions $V_i(l_i)$ and  
$V_j(l_j)$ as well as the parameters $l_i$ and $l_j$,  read
 \begin{equation}
|D_{ij}| \sim \left\{ \begin{array}{llll}
|V_i(l_i)/V_j(l_j)|^{1/2}  \\
|V_i(l_i)|\,l_i\, |V_j(l_j)|^{-1/2}   \\
|V_i(l_i)|^{1/2}\,l_j    \\
|V_i(l_i)|\,l_i\, l_j  
\end{array} \right. \mbox{for}~
\left\{ \begin{array}{llll}
V_i \in {\cal G}'_i ,~V_j \in {\cal G}'_j,\\
V_i \in {\cal G}_i ,~V_j \in {\cal G}'_j,\\
V_i \in {\cal G}'_i ,~V_j \in {\cal G}_j,\\
V_i \in {\cal G}_i ,~V_j \in {\cal G}_j, \end{array} \right. 
\label{20}
\end{equation}
in series (\ref{10}) for $\lambda_{11}$ and  
\begin{equation}
|D_{ji}| \sim  \left\{ \begin{array}{llll}
|V_j(l_j)/V_i(l_i)|^{1/2}   \\
|V_j(l_j)|^{1/2}\, l_i   \\
|V_j(l_j)|\, l_j\, |V_i(l_i)|^{-1/2}    \\
|V_j(l_j)|\, l_i \,l_j   
\end{array} \right. \mbox{for}~
\left\{ \begin{array}{llll}
 V_i \in {\cal G}'_i ,~V_j \in {\cal G}'_j, \\
V_i \in {\cal G}_i ,~V_j \in {\cal G}'_j, \\
V_i \in {\cal G}'_i ,~V_j \in {\cal G}_j,\\
V_i \in {\cal G}_i ,~V_j \in {\cal G}_j, \end{array} \right.
\label{21}
\end{equation}
in series (\ref{13})  for $\lambda_{22}$.
Both the terms  $D_{ij}$ and $ D_{ji}$ ($i < j$) will be finite 
 in the limit as $l_i \to 0$ and $l_j \to 0$ if 
\begin{equation}
\!\!\!\!\!\!\!\!\!\!\!\!\!\!\!\!\!
\left\{  \begin{array}{llllllll}
|V_i(l_i)/V_j(l_j)| \to c >0  \\
|V_i(l_i)|\,l_i\, l_j \to c \ge 0,~ l_i/l_j   \to c \ge 0  \\
 |V_j(l_j)|\,l_i\, l_j \to c \ge 0,~ l_j/l_i   \to c \ge 0   \\
\, l_i/l_j   \to c \ge 0 \\
\, l_j/l_i   \to c \ge 0  \\
|V_i(l_i)|\,l_i\,l_j \to c \ge 0,~|V_j(l_j)|\,l_i\, l_j  \to c \ge 0 \\
 |V_j(l_j)|\,l_i\, l_j  \to c \ge 0    \\
|V_i(l_i)|\,l_i\,l_j \to c \ge 0  
 \end{array} \right. \mbox{for}~ \left\{ \begin{array}{llllllll}
 V_i \in {\cal G}'_i\, ,~V_j \in {\cal G}'_j\,, \\
V_i \in {\cal G}_i \setminus {\cal G}_{i}^0\,,~V_j \in {\cal G}'_j \,, \\
 V_i \in {\cal G}'_i\, ,~V_j \in {\cal G}_j\setminus {\cal G}_j^0\, \\
V_i \in {\cal G}_i^0\,,~V_j \in {\cal G}'_j \,,\\ 
V_i \in {\cal G}'_i\, ,~~V_j \in {\cal G}_j^0\, , \\
V_i \in {\cal G}_i\setminus {\cal G}_i^0\, ,
 ~V_j \in {\cal G}_j\setminus {\cal G}_j^0\,,\\
V_i \in {\cal G}_i^0 \,,
 ~V_j \in {\cal G}_j\setminus {\cal G}_j^0\,,\\
V_i \in {\cal G}_i\setminus {\cal G}_i^0\,,
 ~~V_j \in  {\cal G}_j^0\,. \end{array} \right. 
\label{22}
\end{equation}
These finite limits can be fulfilled if  $l_i$ and $l_j$ 
tend to zero simultaneously,  forming a pencil of paths 
in the $N$-dimensional orthogonal angle with the vertex at $l=0$. 
Therefore we need to describe  these paths projected 
 onto each $(l_i , l_j)$-face. Thus, let us introduce the following pencil
 projections onto a given $(l_i, l_j)$-face: 
\begin{equation}
\begin{array}{lll}
\Gamma_{ij} \!\!&:=&\!\! \left\{(l_i,l_j) \to 0~ |~  l_i / l_j  \to    0 \right\}, \\
\Gamma'_{ij} \!\!&:=&\!\! \left\{(l_i,l_j) \to 0 ~|~  l_i / l_j  \to c >  0 \right\}
= \Gamma'_{ji}\,, \\
\Gamma_{V_ij} \!\!&:=& \!\!\left\{V_i(l_i) \in {\cal G}_i \setminus {\cal G}_i^0\,,
 ~~(l_i,  l_j) \in \Gamma_{ij}~ \big|~  |V_i(l_i)|\,l_i\, l_j \to  0  \right\},\\
 \Gamma'_{V_ij} \!\!&:=& \!\!\left\{V_i(l_i) \in {\cal G}_i \setminus {\cal G}_i^0\,,
 ~(l_i,  l_j) \in \Gamma_{ij} ~ \big|~  |V_i(l_i)|\,l_i\, l_j  \to c > 0  \right\},
\label{23} \end{array}
\end{equation} 
where $i,j = 1, \ldots , N$ and $i \neq j$. Here, the  linear $(l_i , l_j)$-projective
  paths $\Gamma'_{ij} = \Gamma'_{ji}$ determine the  pencil composed of all the straight 
  lines filling in the interior of the $N$-dimensional orthogonal angle. They   
   are  boundary (limit) cluster sets of  both the 
  pencils $\Gamma_{ij}$ and $\Gamma_{ji}$, which consist of curved lines in this interior.
The projective nonlinear paths $\Gamma_{V_ij}$ and  $\Gamma'_{V_ij}$ may be termed as
`adjoint' ($l_i,l_j$)-curves of the strength $V_i(l_i)$. More precisely, 
the paths $\Gamma_{V_ij}$ mean that 
  $l_i/l_j$ tends to zero more slowly than $|V_i(l_i)|\,l_i^2$ (notice that $V_i \in 
  {\cal G}_i \setminus {\cal G}_i^0$), i.e., 
$|V_i(l_i)|\,l_i^2 /( l_i/l_j) \to 0 $, while for the limit sets  $\Gamma'_{V_ij}$,
this ratio is non-zero: $l_i/l_j \sim |V_i(l_i)|\,l_i^2 \to 0$. The unions
\begin{equation}
\begin{array}{ll}
\bar\Gamma_{ij} := \Gamma_{ij} \cup \Gamma'_{ij} = 
\left\{(l_i,l_j) \to 0 ~|~  l_i / l_j  \to c \ge   0 \right\}, \\
  \bar\Gamma_{V_ij} := \Gamma_{V_ij} \cup\Gamma'_{V_ij} =
  \left\{(l_i,  l_j)\to 0 ~ \big|~ l_i/l_j \to 0,\,  
  |V_i(l_i)|\,l_i\, l_j \to c \ge 0  \right\},
 \end{array}
\label{24}
\end{equation}
where $\bar\Gamma_{V_ij} \subset \Gamma_{ij}$\,,   will also be used from now on.
  The full family of the $(l_i , l_j)$-projective
  paths, under which all conditions (\ref{22}) hold true, can be collected 
together  as follows 
\begin{equation}
   \left\{ \begin{array}{lllllllll}
   \Gamma'_{ij} \\
\bar\Gamma_{V_ij} \cup \Gamma'_{ij}\\  \bar\Gamma_{ij} \\ 
 \bar\Gamma_{V_ji} \cup \Gamma'_{ji} \\  \bar\Gamma_{ji} \\
 \bar\Gamma_{V_ij} \cup \bar\Gamma_{V_ji}\cup \Gamma'_{ij} \\ 
   \bar\Gamma_{V_ji} \cup \bar\Gamma_{ij} \\
   \bar\Gamma_{V_ij}\cup \bar\Gamma_{ji}  \\  \bar\Gamma_{ij} \cup \bar\Gamma_{ji} 
\end{array}  \right. \mbox{if} ~~ \left\{
\begin{array}{lllllllll}
V_i \in {\cal G}'_i \,,~V_j \in {\cal G}'_j \,,\\
V_i \in {\cal G}_i \setminus {\cal G}_i^0 \,,~V_j \in {\cal G}'_j \,,\\
V_i \in {\cal G}^0_i \,,~V_j \in {\cal G}'_j\,, \\
V_i \in {\cal G}'_i \,,~V_j \in {\cal G}_j \setminus {\cal G}_j^0 \,,\\
V_i \in {\cal G}'_i \,,~V_j \in {\cal G}^0_j\,, \\
V_i \in {\cal G}_i \setminus {\cal G}_i^0 \,,~V_j \in {\cal G}_j \setminus {\cal G}_j^0 \,,\\
V_i \in  {\cal G}_i^0 \,,~V_j \in {\cal G}_j \setminus {\cal G}_j^0 \,,\\
V_i \in {\cal G}_i \setminus {\cal G}_i^0 \,,~~V_j \in  {\cal G}_j^0 \,,\\
V_i \in  {\cal G}_i^0 \,,~V_j \in  {\cal G}_j^0 \,,
\end{array} \right.~~ (i<j). 
\label{25}
\end{equation}

Thus, under  any of nine conditions (\ref{25})  fulfilled on all 
the $(l_i,l_j)$-faces, the $l \to 0$ limit of the $\Lambda$-matrix elements
$\lambda_{11}$ and $\lambda_{22}$ is finite.  Next, under these conditions,
we also have  $\lambda_{12} \to 0$ and this immediately follows from formulas
(\ref{9}) and (\ref{11}) because $q_i^{-1}t_i$'s and therefore 
all the terms in the series $Q_{12}$ tend to zero.
 Contrary, in expressions (\ref{9}) and (\ref{12}), at least one of the summands 
 $q_it_i $ and possibly some terms in the series $Q_{21}$ 
 in general diverge as $l \to 0$, except for 
some cases as all  appearing  divergences will mutually canceled out and this 
situation will be analyzed below. Hence, 
under the conditions ensuring the finite squeezed limits of $\lambda_{11}$ and
$\lambda_{22}$, the resulting point interactions are {\it separated}.  
Thus, the above results can be summarized in the form of   

{\bf Conclusion 2}\,:  {\it Let all the functions $V_i(l_i)$, $i=1, \ldots , N$, 
 belong to any of the subsets
$ {\cal G}_{i}^0$, $ {\cal G}_{i}\setminus {\cal G}_{i}^0$ or $ {\cal G}'_{i}$
of the  $ \bar{\cal G}_{i}$-set defined by conditions (\ref{15}), (\ref{18}) and (\ref{19}). 
Assume that for at least one layer,  the function $V_i(l_i)$ belongs to 
${\cal G}_{i} \setminus {\cal G}_i^0$ or $ {\cal G}'_{i}$\,.  
Then the element $\lambda_{21}$  diverges in the limit as $l \to 0$ and 
 under any of  conditions 
(\ref{25}) fulfilled on all the $(l_i,l_j)$-faces, $i,j =1,\ldots ,N$ 
($i < j$),   the family of one-point interactions is realized in the squeezing limit with
 the  two-sided boundary conditions on the wave function 
$\psi(x)$ being of the Dirichlet type: $\psi(\pm 0)=0$. }

\section{Resonant tunneling one-point interactions}

We have derived  above the conditions, under which the $l \to 0$ limit of the
matrix elements $\lambda_{11}$ and $\lambda_{22}$ is finite as well as the limit 
relation  $\lambda_{12} \to 0$. Here, on the basis of expressions (\ref{9}) and (\ref{12}),
we will analyze the element $\lambda_{21}$, which 
turns out to be the most singular $\Lambda$-matrix element  as $l \to 0$.
The behavior of each term in this element  depends on the paths, 
along which the squeezing limit is implemented.

As shown above, the squeezed  $\delta$-potentials have been materialized  for  all the 
strengths $V_i(l_i) $ obeying the condition $|V_i(l_i)|\,l_i \to c \ge 0$ and the
corresponding  $l \to 0$ paths were arbitrary, including various repeated limits
$l_i \to 0$, $i=1, \ldots ,N$. In this (regular) case, the connection matrix is 
of form (\ref{16}). In the following for this particular 
case, the summands in $\sum_{i=1}^Nq_it_i$ will be 
referred to as {\it regular} terms. Since no additional restrictions on this
behavior were imposed, the transmission through this point potential 
can be referred to as a {\it non-resonant} tunneling. 

The situation crucially changes   if at least one of the terms $q_i$, 
for which $V_i(l_i) \in \bar{\cal G}_i \setminus {\cal G}_{i}^0  $, 
is present in series (\ref{9}) and (\ref{12}). In this (singular) case,  
$|V_i(l_i)|\,l_i \to \infty$ as $l \to 0$ and in spite of the validity of conditions 
(\ref{25}),  the divergent terms in series (\ref{9}) and (\ref{12}) in general 
will be present. We refer these terms to as {\it singular} ones. One can write 
  $\lambda_{21} = \lambda_{21}^c +\lambda_{21}^\infty$, where 
  $\lambda_{21}^c \to \alpha \in \R$ and each term from $\lambda_{21}^\infty$
diverges  as $l \to 0$.
Under some conditions on the functions $V_i(l_i)$ and the paths
converging  to the origin $l=0$, the cancellation of divergent terms in the group 
 $\lambda_{21}^\infty$  can be materialized, resulting in a finite limit value of
  $\lambda_{21}$. However, this is possible if all the divergent terms are of the  
{\it same} order on certain paths. 
Therefore it is necessary to `measure' the divergence rate of 
these terms by the multiplication them by an appropriate factor $L \to 0$, 
chosen in such a way
that {\it all} the resulting products are finite non-zero quantities in the limit as 
$l \to 0$. In other words, a one-to-one correspondence between these quantities and 
the divergent terms must be provided in the squeezed limit. As a result, the equation 
for the cancellation of divergent terms is of the form  
 $\lim_{l \to 0} (\lambda_{21}^\infty L) =0$, being a constraint on
some limit parameters that describe the configuration $V_1, \ldots , V_N$ and 
the pencil sets.   In the following, the conditions on these limit parameters 
  will be referred to as {\it resonance equations} and their solutions
{\it resonance sets} (the notion  introduced by Golovaty and 
Man'ko \cite{gm}).  

It is important to stress that the divergence rate of terms in series (\ref{9}) and 
(\ref{12}) {\it depends} on the paths, on which the resonance equations are derived. 
Consider, for instance, the convergence of a given term $q_it_i$ on a 
$(l_i,l_j)$-face, for which  $V_i \in {\cal G}_i\setminus {\cal G}^0_i$\,.   
Approaching the origin along the paths from the pencil $\Gamma'_{ij}$
($l_i/  l_j \to c >0$), we have  $|q_i t_i|l_j  \sim |q_i^2|l_il_j \to 0$,
while on the pencil $\Gamma'_{V_ij}$ ($l_i/l_j \sim |V_i|\,l_i^2 \to 0$), 
this limit is non-zero: $|q_i^2|l_il_j \to c >0 $. This means that the divergence
of the term $q_i^2l_i$ on the pencil $\Gamma'_{ij}$ is lower than on  
$\Gamma'_{V_ij}$\,. Therefore the factor $L$ must be chosen in such a way
that all the divergent terms, participating in the cancellation of divergences, 
have to be of the same order if this possible.

The limit parameters 
\begin{equation}
 s_i := \lim_{l_i \to \,0}(q_il_i),~~\tau_i := \lim_{l_i \to 0} ( q_il_it_i )= 
s_i \tan s_i\,,~~i = \overline{1,N},
\label{26}
\end{equation}
will be used for each strength $V_i \in {\cal G}'_i$\,. Next, on paths (\ref{23}), we 
also define
\begin{equation}
\chi_{ij} := \lim_{(l_i,\,l_j) \in \Gamma'_{ij} }\!(l_i/l_j)= \chi_{ji}^{-1} 
~~~\mbox{and}~~~
 \eta_{ij}:= \lim_{(l_i,\, l_j) \in \Gamma'_{V_ij}}\!(q_i^2l_i l_j)
 \label{27}
 \end{equation}
for $V_i \in {\cal G}_i$\,, where $ i,j = 1, \ldots , N$  ($i \neq j)$.  
Clearly, if $V_i \in {\cal G}_i^0$, then $\eta_{ij} \equiv 0$.  
Thus, the cancellation of divergences imposes some constraints on the 
configurations $V_1(l_1), \ldots , V_N(l_N)$ and the  path pencils,
which have to be defined properly in the limit as $l \to 0$. 

Let us consider two  simple examples illustrating how the cancellation procedure
does work, namely  the  pairwise configurations (i) 
 $V_i \in {\cal G}'_i $ and $V_j \in {\cal G}'_j$\,,
 and (ii) $V_i \in {\cal G}_i \setminus {\cal G}_i^0$ and $V_j \in {\cal G}'_j$, 
 whereas the rest  of strengths are regular terms. In case (i), the cancellation 
 equation reads $\lambda_{21}^\infty = q_it_i +q_jt_j =0$ and both the divergent terms in this
 equation are of the same rate in the squeezing limit if the $(l_i,l_j)$-paths belong to
 the pencil $\Gamma'_{ij}$\,.  Therefore this equation can be multiplied by either $l_i$ or
 $l_j$. Thus, multiplying it by $l_i$\,, in the limit as $l_i \to 0$ and $l_j \to 0$, 
 we arrive at the resonance equation
\begin{equation}
\tau_i + \chi_{ij}\tau_j =0,
\label{28}
\end{equation} 
written in terms of limit parameters (\ref{26}) and (\ref{27}), which in their turn depend 
only on the asymptotic behavior of the layer parameters $V_i(l_i)$, $V_j(l_j)$, $l_i$ and 
$l_j$.  In particular, on the paths with $l_i =l_j $ and for the configuration 
$V_i(l_i) = - V_j(l_j) $, equation (\ref{28}) reduces to \cite{c-g}
\begin{equation}
\tan s_i = \tanh s_i,~~~s_i = \lim_{l_i \to 0 }|V_i(l_i)|^{1/2}l_i\,.
\label{29}
\end{equation}  
In the multi-layer case described above,
 the solutions $\{ s_{i,n} \}_{n = - \infty}^\infty$ of this equation  form 
the discrete resonance set  of the point interaction that corresponds to  the potential
$V(x)= a\delta'(x) + b\delta(x)$.  At the values $a =  s_{i,n} $\,, this interaction is
partially transparent, whereas beyond these values the transmission is zero.

Similarly, in the other case (ii), it would be possible to multiply the equation
$\lambda_{21}^\infty = q_i^2l_i +q_jt_j =0$, where both the summands diverge as 
$(l_i,l_j) \to 0$ ($V_i \in {\cal G}_i \setminus {\cal G}^0_i$ and $V_j \in {\cal G}'_j$), 
by $l_i$\,. However, in this case, $|q_i^2|\,l_i^2 \to 0$,
while the other term $\chi_{ij}\tau_j$ is non-zero. This means that the 
divergence of the terms $q_i^2l_i$ and $q_jt_j$ is of different order on 
the pencil $\Gamma'_{ij}$\,, where $l_i$ and  $l_j$ approach the origin linearly. 
 Therefore this pencil is not appropriate for the cancellation of divergences.
Instead, let us consider the cancellation of divergences on the pencil $\Gamma'_{V_ij}$
of curved paths, for which $l_i/l_j \sim |V_i(l_i)|\,l_i^2 \to 0$, where  the divergence rate
of both terms is of the same order:  $|q_i^2|\,l_i \sim |V_i(l_i)|\, l_i  \sim l_j^{-1}$
and $|q_jt_j| \sim |V_j(l_j)| \sim l_j^{-1}$. In this case,  the parameter
$L =l_j$ can be used as a multiplier for the comparison of the divergence of 
$q_i^2l_i$ and $q_jt_j$\,. Then the resonance equation is well-defined on the 
$\Gamma'_{V_ij}$-paths and it   
can be written in terms of finite quantities as follows 
\begin{equation}
\lim_{(l_i,l_j) \in \Gamma'_{V_ij}}\!\!(q_i^2l_i + q_jt_j)\,l_j =
\eta_{ij} + \tau_j =0.
\label{30}
\end{equation}
For a negative strength $V_j$ ($s_j >0$),  this equation (with respect to the variable 
$s_j$) admits a countable set of solutions. The configuration with some $V_i \in {\cal G}_i^0$ 
can also be considered and then in equation (\ref{30}) we have $\eta_{ij} =0$. 
This situation is the 
limiting case as the divergence vanishes due to the solution of the equation 
$\tau_j =0$, forming  the resonance set consisting of points
  $s_j = n_j\pi$ with integers $n_j$.

 Since, in the limit as $l \to 0$, the dyads $D_{ij}$ are finite  under conditions 
(\ref{25}), for the systems with $N \ge 3$, the divergence rate of
the terms in series (\ref{12}) does not exceed that of $q_it_i$'s. Moreover,
due to the presence of the triads
\begin{equation}
T_{ijk} := D_{ij}q_kt_k = (q_iq_k/q_j)t_it_jt_k~~ ~(1 \le i <j< k \le N)
 \label{31}
\end{equation}
  in the $Q_{21}$-series,  rewritten in the form
\begin{equation}
\!\!\!\!\!\!\!\!\!\!\!\!\!\!\!\!\!\!
{\cal Q}_{21}= \sum_{n=1}^{[(N-1)/2]}(-1)^n \!\! \sum_{1 = i_1 < j_1 < 
\ldots < i_n< j_n < k_n =N} 
 D_{i_1j_1} \ldots \stackrel{\surd}{D}_{i_mj_m}\ldots D_{i_nj_n} T_{i_mj_mk_n} ,~~
\label{32}
\end{equation}
where the products $D_{i_mj_m}q_{k_n}t_{k_n}$ have been replaced by the triads 
$T_{i_mj_mk_n}$ and 
the symbol $\surd$  stands for the absence of an indicated term (here the factor 
${D}_{i_mj_m}$), it is possible that the $l \to 0$ limit of these products will be of type 
$0 \cdot \infty \sim c \ge 0$ if $D_{ij} \to 0$. 
Hence, depending on the paths, there are two possibilities: (i) the triads 
$T_{ijk} $ are divergent and the order
of their divergence is the same as that of the terms $q_it_i$'s and (ii)
$|T_{ijk}| \to c \ge 0$, leading to the cancellation of divergences only in the sum
$\sum_{i=1}^N q_it_i$. Clearly, some terms in the latter sum as well as in 
the series $Q_{21}$ may be finite belonging to the group $\lambda_{21}^c$.

Let us analyze the asymptotic behavior of a $T_{ijk}$-triad, where all the 
strengths $V_i$\,, $V_j$ and $V_k$ belong to the corresponding 
${\cal G}'$-sets. In this situation, we have the following asymptotic relation: 
\begin{equation}
T_{ijk} \sim (l_j/l_il_k)\,\tau_i\tau_j\tau_k/s_j^2 \, \sim \,  
l_j^{-1}\chi_{ji}\, \chi_{jk}\, \tau_i\tau_j\tau_k/s_j^2. 
\label{33}
\end{equation}
Hence, in the linear squeezing limit ($l_i \sim l_j \sim l_k$), $T_{ijk}$ diverges as
$l_i^{-1} \sim l_j^{-1} \sim l_k^{-1}$. 
Therefore, if one would like to have the limit $|T_{ijk}| \to 
c \ge 0$, it is necessary that at least  $l_j/l_i \to 0$ or $l_j/l_k \to 0$
sufficiently fast. This nonlinear behavior could be realized on the adjoint paths 
$\Gamma_{V_ji}$ or $\Gamma_{V_jk}$\,, but the strength $V_j$ in this case must belong to
${\cal G}_j \setminus {\cal G}_j^0$, that contradicts with the assumption 
$V_j \in {\cal G}'_j$\,. Therefore the only possibility 
for materializing a finite limit of $T_{ijk}$ is a {\it nonlinear} extension of the
linear squeezing.  To this end, we introduce a parameter $\sigma \in [1, \infty )$ and 
extend the notions of the boundary
pencils $\Gamma'_{ij}$ and $\Gamma'_{V_ij}$ given in equations (\ref{23}) to
\begin{equation}
\begin{array}{ll}
\Gamma^\sigma_{ij} ~\, :=  \left\{ (l_i,\, l_j) \to 0 ~\big|~  
l_i^{1/\sigma} / l_j  \to c >  0 \right\}, \\
 \Gamma^\sigma_{V_ij}  :=  \left\{V_i(l_i) \in {\cal G}_i \setminus {\cal G}_i^0\,,
 ~(l_i, \, l_j) \in \Gamma_{ij} ~ \big|~ |V_i(l_i)|\,l_i\,l_j^{1/\sigma}  \to  c > 0 
  \right\}, \end{array}
\label{34}
\end{equation} 
so that $\Gamma^\sigma_{ij}|_{\sigma=1} 
\equiv \Gamma'_{ij} = \Gamma'_{ji}$ and $\Gamma^\sigma_{V_ij}|_{\sigma=1} \equiv
 \Gamma'_{V_ij}$\,.  
 Since $\sigma \ge 1$, we have $\Gamma_{ij}^\sigma \subset \bar\Gamma_{ij} $ and 
$\Gamma_{V_ij}^\sigma \subset \bar\Gamma_{V_ij}$\,. Therefore conditions (\ref{25}) are
not violated on paths (\ref{34}). Note that for the paths $\Gamma_{V_ij}^{\sigma}$,
as follows from definition (\ref{34}), we have the relation $l_i/l_j^{1/\sigma} 
\sim |V_i(l_i)|\,l_i^2 \to 0$. Accordingly, for each $\sigma$ we introduce a new set of 
the functions $V_i(l_i)$ as 
\begin{equation}
 {\cal G}^\sigma_i  :=  \left\{ V_i(l_i) \in \bar{\cal G}_i \setminus {\cal G}^0_i
   ~\big|~ \lim_{l_i \to 0}|V_i(l_i)|\,l_i^{1 + 1/\sigma}  = c > 0 \right\},
   ~1 \le \sigma < \infty ,
\label{35}
\end{equation}
where ${\cal G}^\sigma_i \big|_{\sigma =1}  \equiv {\cal G}'_i$\,. Next, 
 for  $\sigma \in (1, \infty)$, we have ${\cal G}^\sigma_i \subset {\cal G}_i 
\setminus {\cal G}^0_i $  and $\lim_{\sigma \to  \infty}{\cal G}^\sigma_i ={\cal G}^0_i$,
so that the pencil ${\cal G}^\sigma_i$ is found between the sets
 ${\cal G}^0_i$ and ${\cal G}'_i$\,.
 
Next, 
   we have to  extend the definition of the limit quantities $\chi_{ij}$ and $\eta_{ij}$ 
given by equations (\ref{27}) for  all values $\sigma \in [1, \infty )$. 
Thus, we define ($i,j = 1, \ldots ,  N$)
\begin{equation}
\!\!\!\!\!\!\!\!\!\!\!
\chi_{ij}^\sigma =  \lim_{(l_i,\, l_j) \in \Gamma^\sigma_{ij} }\!(l_i^{1/\sigma}/l_j), ~
i \neq j,~~\mbox{and}~~
 \eta_{ij}^\sigma:= \lim_{(l_i,\, l_j) \in \Gamma^\sigma_{V_ij}}\!\!(q_i^2l_i l_j^{1/\sigma})
~~\mbox{if}~~V_i \in {\cal G}_i,~~ 
\label{35a}
 \end{equation}
where the definition of the pencils $\Gamma^\sigma_{ij}$ and $\Gamma^\sigma_{V_ij}$ 
is given in equations (\ref{34}). According to notations (\ref{27}), 
 $\chi_{ij}^\sigma |_{\sigma =1} \equiv \chi_{ij} =\chi_{ji}^{-1}$, 
$\eta_{ij}^\sigma |_{\sigma =1} \equiv \eta_{ij}$\, and $\eta_{ij}^\sigma \equiv 0$ if
$V_i \in {\cal G}^0_i$\,. Contrary to the case $\sigma =1$,
for $\sigma >1$ it is possible to consider the particular case  $j =i$. From definition 
(\ref{35}), we have $|V_i| \sim l_i^{-(1 +1/\sigma)}$ for $V_i \in {\cal G}^\sigma_i$\,, 
$i = 1, \ldots , N$.
  
Thus, asymptotic relation (\ref{33}) can be generalized to  
 \begin{equation}
 T_{ijk} \sim (l_j^{2/\sigma}/l_il_k)\,l_j^{1 -2/\sigma} \tau_i \tau_j \tau_k /s_j^2 \,\sim \, 
 l_j^{1 -2/\sigma}\chi_{ji}^\sigma\, \chi_{jk}^\sigma\, \tau_i\tau_j\tau_k/s_j^2 \,, 
\label{36} 
 \end{equation}
being valid on the paths, where $l_j^{1/\sigma} \sim l_i$ and  $l_j^{1/\sigma} \sim l_k$\,,
i.e., on the pencil projections $\Gamma'_{ik}$\,, $\Gamma_{ji}^\sigma$ and 
$\Gamma_{jk}^\sigma$\,. Hence,  representation (\ref{36}) coincides with 
(\ref{33}) at $\sigma =1$ and at the same time it also holds true 
 on the whole interval $2 \le \sigma <\infty$. Exactly, at $\sigma =2$, it was possible
to prove the existence of a bound state for the pure $\delta'$-potential
realized on the resonance set \cite{zz21jpa}.

In addition to (\ref{36}), where $V_j \in {\cal G}'_j$\,, 
one can write the asymptotic representation for
the other `lateral' configurations  $V_i \in {\cal G}_i$ and $V_k \in {\cal G}_k$ 
as follows 
\begin{equation}
T_{ijk} \sim l_j^{1 -2/\sigma}\left\{ \begin{array}{lll} 
\par\smallskip \chi_{jk}^\sigma \, \eta_{ij}^\sigma \,
\tau_j\tau_k/s_j^2  \\
\chi_{ji}^\sigma \, \eta_{kj}^\sigma \, \tau_i \tau_j/s_j^2  \par\smallskip  \\ 
\eta_{ij}^\sigma \, \eta_{kj}^\sigma \, \tau_j/s_j^2 \end{array}  \right.
~\mbox{for}~\left\{ \begin{array}{lll} \par\smallskip 
V_i \in {\cal G}_i \setminus {\cal G}^0_i\,,~V_k \in {\cal G}'_k  \,, \\
V_i \in {\cal G}'_i\,,~V_k \in {\cal G}_k \setminus {\cal G}_k^0 \,, \par\smallskip  \\
V_i \in {\cal G}_i \setminus {\cal G}^0_i\,,~V_k \in {\cal G}_k \setminus {\cal G}_k^0
\end{array} \right. 
\label{37}
\end{equation}
and 
\begin{equation}
T_{ijk} \sim l_j^{1 -1/\sigma}\left\{ \begin{array}{lllll} 
\par\smallskip \chi_{jk}^\sigma \, \tau_j\tau_k/s_j^2 \\
\eta_{kj}^\sigma \, \tau_j/s_j^2  \par\smallskip  \\
\chi_{ji}^\sigma \,  \tau_i \tau_j/s_j^2 \par\smallskip  \\ 
 \eta_{ij}^\sigma \, \tau_j/s_j^2  \par\smallskip  \\ 
 \, 0 \end{array}  \right.
~\mbox{for}~\left\{ \begin{array}{lllll} \par\smallskip 
V_i \in  {\cal G}^0_i\,,~V_k \in {\cal G}'_k \,, \\
V_i \in {\cal G}^0_i\,,~V_k \in {\cal G}_k \setminus {\cal G}_k^0 \,, \par\smallskip  \\
V_i \in {\cal G}'_i\,,~V_k \in {\cal G}_k^0 \,, \par\smallskip  \\
V_i \in {\cal G}_i \setminus {\cal G}^0_i\,,~V_k \in  {\cal G}_k^0 \,, \par\smallskip  \\
V_i \in  {\cal G}^0_i\,,~V_k \in  {\cal G}_k^0\,.
\end{array} \right. 
\label{38}
\end{equation}
Equations (\ref{33}) and (\ref{36})--(\ref{38}) can also be used for the case as 
$V_j \in {\cal G}_j$, just setting in these formulas $\tau_j/s^2_j =1$. 
The asymptotic representation given by relations (\ref{37})  holds true
for  $\sigma =1$ and $\sigma \in [2,\infty )$, whereas representation (\ref{38}) is valid 
on the whole interval $1 \le \sigma < \infty$. Below the cases $\sigma =1$ and 
$1 < \sigma < \infty$ will be considered separately.

\subsection{Conditions for the existence of resonance sets 
on the paths with $\sigma =1$ 
} 

The cancellation procedure can particularly be performed on the paths with $\sigma =1$,
which are defined by the pencil projections $\Gamma'_{ij}$ and $\Gamma'_{V_ij}$
[see relations (\ref{23})]. Consider first the case $N=3$, where series  (\ref{32})  
reduces to the single summand $Q_{21} = T_{ijk}$ ($i=1$, $j=2$, $k=3$). Assume that  all the strengths belong to
either ${\cal G}'$- or ${\cal G} \setminus {\cal G}^0$-sets. Then, according to the asymptotic 
representation given by relations (\ref{36}) and (\ref{37}) with $\sigma =1$, the
$T_{ijk}$-triad diverges as $l_j^{-1} \sim l_i^{-1} \sim l_k^{-1}$.
Hence, all the terms in $\lambda_{21}$ are divergent  and the condition for the 
 cancellation of divergences yields the equation
\begin{equation}
q_it_i + q_jt_j +q_kt_k = T_{ijk} \, .
\label{39}
\end{equation} 
For the comparison of the divergence rate of all the terms  on the paths 
with $\sigma =1$, this equation can be multiplied by one of the widths $l_i$, $l_j$
or $l_k$\,, but $V_i$, $V_j$ or $V_k$ must belong to the corresponding
${\cal G}'$-sets. More precisely, if for instance,
the strength  $V_j$ belongs to ${\cal G}_j \setminus {\cal G}_j^0$, resulting in 
the behavior $|q_jt_j| \sim |q_j^2|\,l_j \to \infty$, the multiplication of this term 
by $l_j $ leads to its disappearance as $l_j \to 0$ because $|q_j|\,l_j \to 0 $.
  Therefore the width $l_j$ cannot be used here for `measuring' the
rate of the divergence of $|q_j^2|\,l_j$ and in this case another parameter, say 
$l_i$ or $l_k$ should be chosen as a multiplier if $V_i \in {\cal G}'_i$ or 
$V_k \in {\cal G}'_k$, respectively.  
Thus, at least the strength in one of the summands of the left-hand side of equation 
(\ref{39}), where all the terms are singular, 
has to belong to the corresponding ${\cal G}'$-set. 

Consider now the situation as $V_i \in {\cal G}'_i $ and
$V_j \in {\cal G}_j \setminus {\cal G}_j^0$. According to the above arguments, in this case,
equation (\ref{39}) can be multiplied by $l_i$\,. For $V_k \in {\cal G}'_k$\,,
the resulting products make sense on the paths, where $l_j/l_i \sim |V_j|\,l_j^2 \to 0$,
$l_i \sim l_k$ and $l_j \sim l_k$ [see representation (\ref{33})].  
However, the last two relations contradict with the 
first one. Similarly, for $V_k \in {\cal G}_k \setminus {\cal G}_k^0$\,, we have 
$l_j/l_i \sim |V_j|\,l_j^2 \to 0$, $l_k/l_i \sim |V_k|\,l_k^2 \to 0$ and 
$l_k/l_j \sim |V_k|\,l_k^2 \to 0$.  From the last two relations, one obtains that
$l_i \sim l_j$ and this  again contradicts with the first relation. Thus, the configuration
of the singular strengths $V_i $, $V_j$ and $V_k$ in equation (\ref{39}), where the `middle'
strength $V_j $ belongs to ${\cal G}_j \setminus {\cal G}_j^0$ cannot be used for deriving
resonance equations in the squeezing limit. 
In other words, for the strength configurations with ${\cal G}_j \setminus {\cal G}_j^0$,
it is impossible to find appropriate paths, on which the divergent terms are of the same 
order. However, if $V_j \in {\cal G}'_j$\,, multiplying then equation (\ref{39}) 
by $l_j$ and using asymptotic relations (\ref{33}) and (\ref{37}) with $\sigma =1$, 
one obtains the correct resonance equations, which are collected in table~\ref{tab:table1}. 
\begin{table}
\caption{\label{tab:table1}
 Resonance equations for a three-layer structure  defined on  
 the corresponding  paths  in the case  as   $V_j \in {\cal G}'_j$\,. 
 }
\begin{tabular}{lll}
\hline 
\hline
$V_i(l_i),~V_k(l_k)$ & Resonance equations & Paths \\ 
\hline  \\ \par\smallskip 
$V_i \in {\cal G}'_i\,,~V_k \in {\cal G}'_k$  
& $\chi_{ji}\tau_i +  \tau_j + \chi_{jk}\tau_k =
\chi_{ji}\chi_{jk} \tau_i \tau_j \tau_k/ s^2_j $ & 
$\Gamma'_{ij}\,, \,  \Gamma'_{jk}  $
  \\  \par\smallskip 
$V_i \in {\cal G}_i \setminus {\cal G}^0_i\,,~V_k \in {\cal G}'_k$ 
 & $\eta_{ij} + \tau_j +\chi_{jk}\tau_k = \eta_{ij} \chi_{jk} \tau_j \tau_k/s_j^2$  
&  $\Gamma'_{V_ij}\,,  \,   \Gamma'_{jk}  $ 
    \\  \par\smallskip 
$V_i \in {\cal G}'_i\,,~V_k \in {\cal G}_k \setminus {\cal G}_k^0$  & 
  $\chi_{ji}\tau_i +  \tau_j + \eta_{kj}= 
\eta_{kj} \chi_{ji} \tau_i\tau_j/ s_j^2$  & 
$\Gamma'_{ij}\,,   \,  \Gamma'_{V_kj}  $
     \\  \par\smallskip 
$V_i \in {\cal G}_i \setminus {\cal G}^0_i\,,~V_k \in {\cal G}_k \setminus {\cal G}_k^0$ &  
$\eta_{ij}  + \tau_j + \eta_{kj} =\eta_{ij} \,  \eta_{kj}\tau_j/s_j^2$   & 
 $\Gamma'_{V_ij}\,, \,  \Gamma'_{V_kj}  $ 
 \\   
\hline
\hline
\end{tabular}
\end{table}
    Here, all the four terms in equation 
 (\ref{39}) are divergent being of the same order, so that all the terms in the 
 resonance equations are finite  and  non-zero. 
Note that for the configurations pointed out in the second, the third and the fourth
lines in table~\ref{tab:table1}, it is impossible to derive resonance equations 
on the pencils $\{\Gamma'_{ij}\,, \, \Gamma'_{jk}\}$ because the divergent terms of
equation (\ref{39}) in this situation are of different order.

Next, we have to consider the configurations $V_i$, $V_j$ and $V_k$ ($N=3$), where 
one or two strengths belongs to ${\cal G}^0$-sets. Thus,
 setting in table~\ref{tab:table1} $\tau_j =0$ and $\tau_j/s^2_j =1$, one obtains 
the resonance equations for the configuration with  $V_j \in {\cal G}^0_j$\,. 
The general form of these equations reads
\begin{equation}
A_{ij} +  A_{kj} = A_{ij}A_{kj}\,,
\label{40}
\end{equation}
where ($i \neq j$)
\begin{equation}
A_{ij} := \left\{ \begin{array}{ll} \lim_{(l_i,\,l_j)  \in \Gamma'_{ij} }(q_it_il_j)=  
 \chi_{ji}\tau_i & \mbox{if}~~V_i \in {\cal G}'_i \,,  \\
\lim_{(l_i,\,l_j)  \in \Gamma'_{V_ij} } (q_i^2l_il_j)= 
 \eta_{ij} & \mbox{if}~~V_i \in {\cal G}_i \setminus {\cal G}_{i}^0\,. 
   \end{array} \right.
 \label{41} 
\end{equation}
The solution of equations (\ref{40})  are curves  on
the $\{ A_{ij}, A_{kj} \}$-plane. The corresponding resonance sets 
 have been analyzed in detail 
for various situations earlier  in works \cite{z18aop,zz21jpa}. 
Thus, we conclude that equation (\ref{39})
with the singular lateral terms $q_it_i$ and $q_kt_k$ multiplied by 
$l_j$ makes sense only if the middle strength $V_j \in {\cal G}'_j \cup {\cal G}_j^0$. 

Finally, for the case $N=3$, where one of the lateral strengths $V_i$ or $V_k$ 
(with the $\delta$-like shrinking) belongs  to 
${\cal G}^0_{i}$ or ${\cal G}^0_k$, respectively, we have to use asymptotic
representation (\ref{38}), where $|T_{ijk}| \to c \ge 0$. 
For instance, for the structure with $V_i \in {\cal G}'_i$\,,
$V_j \in {\cal G}'_j$ and $V_k \in {\cal G}^0_k$\,, multiplying equation (\ref{39}) by $l_i$\,,
we arrive at equation (\ref{28}). Similarly, for $V_i \in {\cal G}'_i$\,,
$V_j \in {\cal G}_j \setminus {\cal G}_j^0$ and $V_k \in {\cal G}^0_k$\,,
one obtains  $\tau_i + \eta_{ji} =0$, in fact, equation (\ref{30}). In both these situations,
the cancellation of divergences occurs only between two terms in equation (\ref{39})
as in a double-layer system.

The above arguments can be extended to the general case with $N > 3$. 
Consider the configuration $V_1(l_1), \ldots , V_N(l_N)$, for which 
at least three terms $q_it_i$,  $q_jt_j$ and $q_kt_k$ ($1 \le i < j <k \le N$) 
 diverge in the squeezing limit. Then, according to asymptotic relations 
 (\ref{36}) and (\ref{37}), the corresponding triad $T_{ijk}$ will also be 
a divergent term at $\sigma =1$. Let the lateral strengths $V_i$ and $V_k$ 
belong to the corresponding
${\cal G}'$- or ${\cal G}\setminus {\cal G}^0$-sets, while 
$V_j \in {\cal G}_j\setminus {\cal G}_j^0$. For the comparison of the divergence rate
of these four terms, one can multiply them by some $l_m$\,,
$m = 1, \ldots , N$ ($m \neq i, j, k$).
Thus, multiplying the sum $q_it_i + q_j^2l_j + q_kt_k$ ($V_i \in {\cal G}'_i$ and 
$V_k \in {\cal G}'_k$) and $T_{ijk} = q_it_i\, l_j\, q_kt_k$ by $l_m$, one obtains that
the resulting products will be finite and non-zero quantities on the paths characterized 
by the relations $l_i \sim l_m$, $l_j/l_m \sim |V_j|\,l_j^2 \to 0$, $l_k \sim l_m$ and 
$l_j \sim l_k$\,. From the last two relations we have $l_j \sim l_m$, but this contradicts 
with $l_j/l_m \to 0$. Next, for instance, in the case with 
$V_i \in {\cal G}_i\setminus {\cal G}_i^0$ and 
$V_k \in {\cal G}_k\setminus {\cal G}_k^0$, one obtains the relations 
$l_i/l_m \sim |V_i|\,l_i^2 \to 0$, $l_j/l_m \sim |V_j|\,l_j^2 \to 0$,
$l_k/l_m \sim |V_k|\,l_k^2 \to 0$ and $l_k/l_j \sim |V_k|\,l_k^2 \to 0$.
Again, from the last two relations we obtain  $l_j \sim l_m$ that contradicts with
$l_j/l_m \to 0$. This zero convergence is replaced by
 $l_j \sim l_m$  if $V_j \in {\cal G}'_j$\,, 
so that the contradiction vanishes. In the case as $V_j \in {\cal G}_j^0$\,, the 
restriction $l_j/l_m \to 0$ is absent. Hence, the `internal' strengths  
$V_j \in {\cal G}_j\setminus {\cal G}_j^0$ with $j=2, \ldots , N-1$ have to be excluded 
in the procedure of deriving resonance equations and, as a result, 
 each strength $V_j$ in the sum
$\sum_{i=2}^{N-1}q_it_i$ must belong to either ${\cal G}'_j$ or ${\cal G}^0_j$. 
Note that the lateral terms  in the sum $\sum_{i=1}^{N}q_it_i$ (from one to several ones)
may belong to ${\cal G}^0$-sets, so that the next neighboring strengths appear
to be singular ones. Then, the applying of relations (\ref{38}) results in the same 
result: for the existence of resonance equations,
 the internal strengths must belong to ${\cal G}'\cup{\cal G}^0$-sets.

Thus, multiplying  the element $\lambda_{21}$ given by series (\ref{9}) and (\ref{12}) 
 by one of $l_j$ with $j =2, \ldots , N-1$, from the condition  $\lim_{l \to 0}
 (\lambda_{21}^\infty \, l_j) = \lim_{l \to 0}(\lambda_{21} l_j) =0$,
 we obtain  the limit equation 
 \begin{equation}
 \!\!\!\!\!\!\!\!\!\!\!\!\!\!\!\!\!\!\!\!\!\!\!
 \sum_{i=1}^N A_{ij} \, + \!\!
  \sum_{n=1}^{[(N-1)/2]}(-1)^n \!\!\!\!\!\!\!\!\!\!\!\!\!\!\!
  \sum_{1 = i_1 < j_1 <i_2 <j_2   < \ldots < i_n< j_n < k_n =N} 
  \!\!\! \!\!\!\!\!\!\!\!\!\!\!\!\!\!\!\!\!\!\!
  ( A_{i_1j_1}\tau_{j_1}/s^2_{j_1}) \ldots (A_{i_nj_n}\tau_{j_n}/s^2_{j_n}) A_{k_nj}=0, ~~
\label{42} 
\end{equation}
where  we have used the limits
\begin{equation}
 \left\{ \begin{array}{llll}   \lim_{(l_i,\,l_j) \in \Gamma'_{ij} }D_{ij}   \\
\lim_{(l_i,\,l_j) \in \Gamma'_{ij} }D_{ij}   \\
\lim_{(l_i,\,l_j) \in \Gamma'_{V_ij} }D_{ij}  \\
 \lim_{(l_i,\,l_j) \in \Gamma'_{V_ij} }D_{ij}
  \end{array} \right\} = A_{ij}\tau_j/s_j^2 ~~  \mbox{if}~~\left\{ \begin{array}{llll}
  V_i \in {\cal G}'_i\,,~  V_j \in {\cal G}'_j\,,\\
V_i \in {\cal G}'_i\,, ~V_j \in {\cal G}_j\,, \\
V_i \in {\cal G}_i\setminus {\cal G}_i^0\,,
 \, V_j \in {\cal G}'_j\, , \\
 V_i \in {\cal G}_i  \setminus {\cal G}_i^0\,, \, V_j \in {\cal G}_j\, .
  \end{array} \right.
 \label{43} 
\end{equation}
Here, the limit quantities $A_{ij}$ are defined by equations (\ref{41}). In addition 
to these equations, notice that $A_{ij}=0$ if $V_i \in {\cal G}^0_i$ ($i \neq j$). 
Next, in equation (\ref{42}),  
 $A_{jj} =  \tau_j $ if $V_j \in {\cal G}'_j$\,, while
 $A_{jj}  =0$ and $\tau_j/ s^2_j =1$ if $V_j \in {\cal G}_j^0$. 
Clearly, all the terms from the group $\lambda^c_{21}$ disappear in equation (\ref{42}). 
All the equations obtained from the condition  $\lim_{l \to 0}(\lambda_{21} l_j) =0$,
where $j=2, \ldots , N-1$, are equivalent, so that the resonance sets do not depend on
the subscript $j$ in equation (\ref{42}).
 
Thus, the solutions  of  resonance equation  (\ref{42}) determine 
the conditions on the limit parameters
 $\chi_{ij}$, $s_i$ and $\eta_{ij}$ under which a non-zero (resonant)
transmission through the squeezed structure is possible. These conditions
 form one or several hypersurfaces in
the  $\{ \ldots , \chi_{ji}, \ldots , s_i , \ldots , \eta_{ij}, \ldots \}$-space.
The number and  structure of resonance sets (hypersurfaces) depends on the configuration 
$V_1(l_1), \ldots , V_N(l_N)$ and the corresponding paths converging to the origin $l = 0$. 
For a fixed number of layers $N$,
several resonance sets can exist and its number  increases with  $N$. 

Finally, consider  the particular case as all the functions $V_i(l_i)$ are non-negative.
For realizing a non-trivial point interaction in the singular 
squeezing limit, assume that some 
of these functions $V_i(l_i) \in \bar{\cal G}_i \setminus {\cal G}_i^0$
 are positive (barriers).  Then  $s_i \in \I $,
 so that $\tau_i <0$ ($s^2_i <0$) and $\eta_{ij}< 0$. As a result, 
 all the terms  in equation (\ref{42}) become  negative, 
 and consequently,  the total left-hand expression
in this equation will also be  negative. Hence, in the case of the absence of 
well-like potential strengths $V_i$\,, there are no 
cancellations of divergences  resulting in a finite limit of  $\lambda_{21}$\,. 

However, in the case if at least one well-like potential strength ($V_i <0$) is present,   
equation (\ref{42}) admits a countable set of non-trivial solutions due to the functions 
$\tan s_i$ with $s_i \in \R$. This leads to 
 the realization of non-separated point interactions 
in the squeezing limit. For instance, one of the particular cases illustrating  the existence 
of solutions to equation (\ref{42}) is the configuration, where each divergent potential
strength $V_i$ belongs to the set ${\cal G}'_i$ and all these potential  strengths 
are defined on the  paths $\Gamma'_{ij}$. In this case, all the divergent terms
in (\ref{42}) are of the same order. Hence, in the presence of at least one well-like
potential strength, the cancellation of divergences provides the appearance of 
non-separated point interactions in the squeezing limit.  Thus, from  the above 
arguments we single out the results, which can be formulated as

{\bf Conclusion 3}\,: {\it (a) The resonance sets for $\sigma =1$
are obtained as solutions to limit equation (\ref{42}) defined for the configurations 
$V_1(l_1), \ldots , V_N(l_N)$, $i=1, \ldots , N$, 
where the lateral  strengths $V_i $ ($i =1,N$) are singular, i.e., they belong 
to either the ${\cal G}'$- or ${\cal G}\setminus {\cal G}^0$-sets, whereas 
 the internal strengths  $V_i$ ($i= 2, \ldots , N-1$) are from the 
 ${\cal G}'_i \cup {\cal G}_i^0$-sets. (b)
 In the case as some lateral strengths (from one to several) 
belong to the corresponding ${\cal G}^0$-sets, more precisely,  
$V_1 \in {\cal G}^0_1\,, \ldots ,  V_p \in {\cal G}^0_p$ (left layers) and
$V_{N-r} \in {\cal G}^0_{N-r}\,, \ldots ,   V_N \in {\cal G}^0_N$  (right layers),
so that the strengths $V_{p+1}$ and $V_{N-r-1}$ are singular, now 
the strengths $V_{p+2}\,, \ldots , V_{N-r-2}$ must belong to either the ${\cal G}'$-
or ${\cal G}^0$-sets.  If one of the lateral strengths or both these 
are regular, i.e., $V_1 \in {\cal G}_1^0$ and $V_N \in {\cal G}_N^0$, then 
only the $\delta$-like squeezing of these strengths makes sense. 
 (c) The presence in the configuration $V_1\,, \ldots , V_N$ at least 
 one  strength of a well-like form, say,  $V_i <0$, implies 
the existence of a countable set of solutions to the resonance equations. }

\subsection{Conditions for the existence of resonance sets in the case  $\sigma >1$  }

Similarly to the case  with $\sigma =1$, consider first the situation as $N=3$ and analyze 
in detail all possible 
cancellations of divergences in $\lambda_{21}$ on the paths with $\sigma >1$. 
Here, the cancellation 
can also be possible within equation (\ref{39}), where $i =1$, $j=2$ and $k =3$.
Assume that all the terms $q_it_i$, $q_jt_j$ and $q_kt_k$ diverge in the squeezing limit.
The configuration $V_i \in {\cal G}'_i$, $V_j \in {\cal G}'_j$ and $V_k \in {\cal G}'_k$
is the most singular case.
The divergence rate of all the terms in this case could be compared by multiplying 
equation (\ref{39}) 
by one of the factors $l^{1/\sigma}_i$, $l^{1/\sigma}_j$ or $l^{1/\sigma}_k$
with $\sigma >1$. 
However, the multiplication by any of these factors does not lead to finite values.
For instance, the resulting term $l_i^{1/\sigma} q_it_i  \sim l_i^{1/\sigma -1}\tau_i$
diverges for $\sigma >1$ as $l_i \to 0$. Hence, in the three-layer structure, 
the cancellation of divergences on the paths with $\sigma >1$ 
is impossible if all the three strengths $V_i$, $V_j$ and $V_k$ belong to the 
corresponding ${\cal G}'$-sets. Therefore at least one of the strengths must belong to
the ${\cal G}$-set, which can be either lateral ($V_i$ or $V_k$) or middle ($V_j$).

In addition to relations (\ref{36}) and (\ref{37}), for the case $V_j \in {\cal G}'_j$\,,
 we also need the asymptotic representation
\begin{equation}
T_{ijk} \sim \left\{ \begin{array}{lll} 
\par\smallskip 
l_i^{-1/\sigma}\chi_{jk} \, \eta_{ii}^\sigma \, \tau_j\tau_k/s_j^2  \\
l_k^{-1/\sigma} \chi_{ji} \, \eta_{kk}^\sigma \, \tau_i \tau_j/s_j^2  \par\smallskip  \\ 
l_i^{-1/\sigma}\eta_{ii}^\sigma \, \eta_{kj} \, \tau_j/s_j^2 \par\smallskip  \\ 
 l_k^{-1/\sigma}\eta_{ij} \, \eta_{kk}^\sigma \, \tau_j/s_j^2 
 \end{array}  \right.
~~\mbox{for}~\left\{ \begin{array}{lll} \par\smallskip 
V_i \in {\cal G}_i^\sigma \,,\,V_k \in {\cal G}'_k  \,, \\
V_i \in {\cal G}'_i\,,\,V_k \in {\cal G}_k^\sigma  \,, \par\smallskip  \\
V_i \in {\cal G}_i^\sigma \,,\,V_k \in {\cal G}_k \,, \par\smallskip  \\
V_i \in {\cal G}_i \,,\,V_k \in {\cal G}_k^\sigma \,.
\end{array} \right. 
\label{45}
\end{equation}
Using  next these relations after the multiplication of equation (\ref{39}) 
by $l_i^{1/\sigma}$ and $l_i^{1/\sigma}$, respectively, we get the four resonance equations
for $V_j \in {\cal G}'_j$ and $\sigma >1$, which are collected  in 
table~\ref{tab:table2}. Here, the cancellation of divergences occurs between 
all the four terms if 
$V_i \in {\cal G}_i \setminus {\cal G}^0_i$ or $V_k \in {\cal G}_k \setminus {\cal G}^0_k$\,.
In the case as $V_i \in {\cal G}^0_i$ and $V_k \in {\cal G}^0_k$, we have accordingly 
$\eta^\sigma_{ii} = 0$ and $\eta^\sigma_{kk} = 0$ in table~\ref{tab:table2} 
and therefore only two terms take place in the cancellation.
\begin{table}
\caption{\label{tab:table2}
 Resonance equations  and available paths for a three-layer structure, resulting from the multiplication of 
   equation   (\ref{39}) by $l_i^{1/\sigma}$ or $l_k^{1/\sigma}$, where 
    $V_j \in {\cal G}'_j$ and $\sigma >1$.  }
\begin{tabular}{lll}
\hline 
\hline
$V_i(l_i)$,~~ $V_k(l_k)$ & Resonance equations & Paths \\ 
\hline  \\ \par\smallskip 
$V_i \in {\cal G}_i^\sigma $\,,  $V_k \in {\cal G}'_k$  
& $ \eta^\sigma_{ii} + \chi^\sigma_{ij}\tau_j + \chi^\sigma_{ik}\tau_k = 
\chi_{jk}\, \eta^\sigma_{ii} \, \tau_j\tau_k/s_j^2$  &  
 $  \Gamma'_{jk}$\,,  \, $ \Gamma^\sigma_{ij}$\,, \, $ \Gamma^\sigma_{ik}$
    \\  \par\smallskip 
$ V_i \in {\cal G}'_i $\,, $V_k \in {\cal G}_k^\sigma $ 
&  $\chi^\sigma_{ki}\tau_i + \chi_{kj}^\sigma \tau_j + 
 \eta_{kk}^\sigma  =\chi_{ji}\, \eta^\sigma_{kk}\, \tau_i\tau_j/s_j^2 $  &  
$\Gamma'_{ij}$\,, \,  $\Gamma^\sigma_{ki} $\,, \,  $\Gamma^\sigma_{kj}$
  \\  \par\smallskip 
$V_i \in {\cal G}_i^\sigma $\,,  $V_k \in {\cal G}_k$
 &  $   \eta_{ii}^\sigma  +\chi_{ij}^\sigma \tau_j + \eta^\sigma_{ki} = 
 \eta^\sigma_{ii} \, \eta_{kj} \, \tau_j/s_j^2  $  & 
 $\Gamma'_{V_kj}$\,, \,   $ \Gamma^\sigma_{ij} $\,, \, $ \Gamma^\sigma_{V_ki}  $    
 \\   \par\smallskip 
$V_i \in {\cal G}_i$\,,  $V_k \in {\cal G}_k^\sigma $
 &  $   \eta_{ik}^\sigma  +\chi_{kj}^\sigma \tau_j + \eta^\sigma_{kk} = 
   \eta_{ij} \, \eta_{kk}^\sigma \, \tau_j/s_j^2  $  & 
 $\Gamma'_{V_ij}$\,, \,   $ \Gamma^\sigma_{kj} $\,, \, $ \Gamma^\sigma_{V_ik}  $ \\    
\hline
\hline
\end{tabular}
\end{table}

In the `middle' case as $V_j \in {\cal G}_j$, we have to apply the asymptotic representation
given by equations (\ref{36}) and (\ref{37}), where $\sigma \ge 2$ resulting in the limit
$|T_{ijk}| \to c \ge 0$. Multiplying equation (\ref{39}) by $l_j^{1/\sigma}$, we obtain 
the four  resonance equations, which  are present in 
table~\ref{tab:table3}.
\begin{table}
\caption{\label{tab:table3}
 Resonance equations  and available paths for a three-layer structure, resulting from the multiplication of 
   equation   (\ref{39}) by $l_j^{1/\sigma}$, where  $V_j \in {\cal G}_j^\sigma $ 
   and $\sigma \ge 2$.
  }
\begin{tabular}{lll}
\hline 
\hline
$V_i(l_i)$,~~~ $V_k(l_k)$ & Resonance equations & Paths \\ 
\hline  \\ \par\smallskip 
$V_i \in {\cal G}'_i$\,,  $V_k \in {\cal G}'_k$  & $ \chi_{ji}^\sigma \tau_i 
+ \eta^\sigma_{jj} + \chi^\sigma_{jk}\tau_k = 0$ &  
$  \Gamma^\sigma_{ji}\,,  \, \Gamma^\sigma_{jk} $ 
\\ \par\smallskip 
$ V_i \in {\cal G}'_i $\,,  $V_k \in {\cal G}_k \setminus {\cal G}^0_k $ 
&  $\chi^\sigma_{ji}\tau_i + \eta_{jj}^\sigma  + \eta_{kj}^\sigma  =0 $  
&   $  \Gamma^\sigma_{ji}\,,  \, \Gamma^\sigma_{V_kj} $ 
  \\  \par\smallskip 
$V_i \in {\cal G}_i \setminus {\cal G}^0_i$\,,  $V_k \in {\cal G}'_k$  
& $ \eta^\sigma_{ij} + \eta^\sigma_{jj} + \chi^\sigma_{jk}\tau_k = 0$  
&   $ \Gamma^\sigma_{V_ij}\,,\, \Gamma^\sigma_{jk} $ 
    \\  \par\smallskip 
$V_i \in {\cal G}_i \setminus {\cal G}^0_i$\,, 
 $V_k \in {\cal G}_k \setminus  {\cal G}^0_k$
 &  $   \eta_{ij}^\sigma  +\eta^\sigma_{jj}  + \eta^\sigma_{kj} = 0 $
   &  $ \Gamma^\sigma_{V_ij}\,,   \, \Gamma^\sigma_{V_kj}  $    
 \\   
\hline
\hline
\end{tabular}
\end{table}
Note that the paths given in this table and those derived for the existence of relations 
(\ref{36}) and (\ref{37}) are the same. The general form of the resonance  equations 
in table~\ref{tab:table3} reads
\begin{equation}
A^\sigma_{ij} +  A^\sigma_{kj} = 0,
\label{46}
\end{equation}
where 
\begin{equation}
A^\sigma_{ij} := \left\{ \begin{array}{ll} \lim_{(l_i,\,l_j)  \in \Gamma^\sigma_{ji} }
(l_j^{1/\sigma}\!q_it_i)=  
 \chi^\sigma_{ji}\tau_i & \mbox{if}~~V_i \in {\cal G}'_i \,, ~ i \neq j ,\\
\lim_{(l_i,\,l_j)  \in \Gamma^\sigma_{V_ij} } (q_i^2l_il_j^{1/\sigma})= 
 \eta^\sigma_{ij} & \mbox{if}~~V_i \in {\cal G}_i \setminus {\cal G}_{i}^0\,. 
   \end{array} \right.
 \label{47} 
\end{equation}
The solution of equations (\ref{46})  are curves  on
the $\{ A_{ij}^\sigma, A_{kj}^\sigma \}$-plane, which depend on the parameter $\sigma \ge 2$. 
The corresponding resonance sets for the structure composed of two separated layers 
($N=3$, $V_j \equiv 0$)  have  been analyzed  in works \cite{z18aop,zz21jpa}.

Thus, we have established in the case $N=3$ that for the existence of resonance equations, 
at least one of the strengths among  $V_i$\,, $V_j$ and $V_k$ must belong to the 
corresponding ${\cal G}$-set. For illustration of this rule for $N >3$, let us write
a couple of resonance equations if $N=4$. For this particular case, 
 series (\ref{32}) reduces to the sum
\begin{equation}
Q_{21} =  T_{ijk} +T_{ijm}+T_{ikm}+T_{jkm} ~~(i=1,\, j=2,\, k=3, \, m=4)
\label{48}
\end{equation}
and the equation for the cancellation of divergences $\lambda_{21}^\infty =0$ becomes
$\sum_{i=1}^4q_it_i = Q_{21}$ if all the terms $q_it_i$'s are divergent.

Let us consider first the configuration as one of the  lateral strengths, say $V_m$\,, 
belongs to  the ${\cal G}_m$-set. Assume that
 $V_i \in {\cal G}'_i$\,, $V_j \in {\cal G}'_j$\,, 
 $V_k \in {\cal G}'_k$ and  $V_m \in {\cal G}_m$\,.
Then the equation for the cancellation of divergences takes the form
\begin{eqnarray}
&& q_it_i + q_jt_j +q_kt_k +q_m^2l_m  \nonumber \\
&=& {q_iq_k \over q_j}t_it_jt_k + {q_iq_m \over q_j}t_it_jt_m +
{q_iq_m \over q_k}t_it_kt_m + {q_jq_m \over q_k}t_jt_kt_m\,. 
\label{49}
\end{eqnarray}
The multiplication of this equation by $l_m^{1/\sigma}$ yields the resonance equation
\begin{eqnarray}
&& \chi_{mi}^\sigma \tau_i + \chi_{mj}^\sigma \tau_j +\chi_{mk}^\sigma \tau_k +
\eta_{mm}^\sigma \nonumber \\
& =& \chi_{ji}\tau_i {\tau_j \over s_j^2}\chi_{mk}^\sigma \tau_k
+ \left(\! \chi_{ji}\tau_i {\tau_j \over s_j^2} +
\chi_{ki}\tau_i {\tau_k \over s_k^2} +
\chi_{kj}\tau_j {\tau_k \over s_k^2}\right) \eta^\sigma_{mm} \,.
\label{50}
\end{eqnarray}
Here, in each of the four triads [see equation (\ref{48})], the middle strength is from
the corresponding ${\cal G}'$-set and therefore representation (\ref{45}) has been applied,
similarly as for deriving the first equation in table~\ref{tab:table2}.  
In  equation (\ref{50}), the available paths are characterized by the  relations
$l_i \sim l_j \sim l_k \sim l_m^{1/\sigma}$ forming 
 the paths $ \Gamma'_{ij}\,,\, \Gamma'_{ik}\,,\, \Gamma'_{jk}\,,\,
\Gamma^\sigma_{mi}\,,\,  \Gamma^\sigma_{mj}\,,\, \Gamma^\sigma_{mk} $ with
$\sigma \in (1,\infty)$. Besides these conditions, an additional restriction appears 
on the strength $V_m$\,, namely, this strength must belong to 
${\cal G}_m^\sigma \subset {\cal G}_m \setminus {\cal G}_m^0$\,. In the case as
$V_m \in {\cal G}_m^0$\,, in equation (\ref{50}) we have $\eta^\sigma_{mm} =0$
and therefore only the four terms participate  in removing divergences, instead of
all the eight terms.
 
Consider now a four-layer system with the configuration as one of the strengths
belonging to a ${\cal G}$-set is localized between the strengths from ${\cal G}'$-sets,
e.g., $V_i \in {\cal G}'_i$\,, $V_j \in {\cal G}_j$\,, $V_i \in {\cal G}'_k$ and 
$V_m \in {\cal G}'_m$\,. Here, due to representation (\ref{36}), 
the squeezing limits of the triads
$T_{ijk} = q_it_i\,l_j\,q_kt_k$ and $T_{ijm} = q_it_i\, l_j \,q_mt_m$ are finite 
on the paths, where 
$l_i \sim l_k \sim l_m \sim l_j^{1/\sigma}$. Therefore these triads
do not participate in the cancellation of divergences and they have to be omitted
because their middle strength $V_j$ belongs to the ${\cal G}_j$-set. As a result,
the divergent terms are contained in the equation
\begin{equation}
 q_it_i + q_j^2l_j +q_kt_k +q_mt_m = 
 {q_iq_m \over q_k}t_it_kt_m + q_j^2l_j {t_k \over q_k}q_m t_m \,.
\label{51}
\end{equation}
Multiplying this equation by $l_j^{1/\sigma}$ and using relations (\ref{45}), we arrive at
the resonance equation
\begin{equation}
 \chi_{ji}^\sigma \tau_i + \eta_{jj}^\sigma +\chi_{jk}^\sigma \tau_k +
\chi_{jm}^\sigma \tau_m  = \chi_{ki}\tau_i {\tau_k \over s_k^2}\chi_{jm}^\sigma \tau_m
+ \eta^\sigma_{jj} {\tau_k \over s_k^2}\chi_{km}\tau_m  \,,
\label{52}
\end{equation}
valid under the conditions  $l_i \sim l_k \sim l_m \sim l_j^{1/\sigma }$, resulting in
 the paths $ \Gamma'_{ik}$\,, $ \Gamma'_{im}$\,, $\Gamma'_{km}$\,,
$\Gamma^\sigma_{ji}$\,,  $\Gamma^\sigma_{jk}$\,, $\Gamma^\sigma_{jm}$\,. 
Since representation (\ref{36}) has been used here, we have to choose  
$\sigma \in [2,\infty)$, instead of the interval $1 < \sigma < \infty$. 

The similar situation takes place as   
 $V_i \in {\cal G}'_i$\,, $V_j \in {\cal G}_j$\,, $V_k \in {\cal G}'_k$ and 
$V_m \in {\cal G}_m  $\,. Here, again the same two triads
$T_{ijk}$ and $T_{ijm}$ have the middle strength from ${\cal G}_j$ and for the other
two triads $T_{ikm}$ and $T_{jkm}$, the middle strength $V_k$ belongs to ${\cal G}'_k$\,.
As a result, again the right-hand part of the  resonance equation consists of two
summands. Indeed,  multiplying the equation 
\begin{equation}
 q_it_i + q_j^2l_j +q_kt_k +q_m^2l_m =  
 q_it_i {t_k \over q_k}q_m^2l_m + q_j^2l_j {t_k \over q_k}q_m^2l_m 
\label{53}
\end{equation}
by $l_j^{1/\sigma}$ and applying the representation given by equations (\ref{36})
and (\ref{37}) to the triads $T_{ijk}$ and $T_{ijm}$\,, and  representation (\ref{45})
to $T_{ikm}$ and $T_{jkm}$\,,   we get the resonance equation in the  form 
\begin{equation}
 \chi_{ji}^\sigma \tau_i + \eta_{jj}^\sigma +\chi_{jk}^\sigma \tau_k +
\eta_{mj}^\sigma   = \chi_{ji}^\sigma \tau_i {\tau_k \over s_k^2}\eta_{mk}
+ \eta^\sigma_{jj} {\tau_k \over s_k^2}\eta_{mk}  \,.
\label{54}
\end{equation}
In this equation, the relations $ l_i \sim l_k \sim l_j^{1/\sigma}$ and 
$|V_m|l_m \sim l_j^{- 1/\sigma}$ take place, so that the available paths, on
which equation (\ref{54}) is well-defined, are 
$ \Gamma'_{ik}$\,, $\Gamma'_{V_mi}$\,, $\Gamma'_{V_mk}$\,, $\Gamma^\sigma_{ji}$\,,
$\Gamma^\sigma_{jk}$\,, $\Gamma^\sigma_{V_mj}$ with $\sigma \in [2, \infty)$.
The form of equations (\ref{52}) and (\ref{54}) resembles a `mixture' of the 
equations collected in tables~\ref{tab:table2} and \ref{tab:table3}.

Consider now the configuration $V_i \in {\cal G}'_i$\,, $V_j \in {\cal G}_j$\,, 
$V_k \in {\cal G}_k$ and $V_m \in {\cal G}_m  $\,. Here, all the triads contain 
the middle strengths $V_j$ and $V_k$\,, belonging to the ${\cal G}_j$- 
and ${\cal G}_k$-sets. Therefore only  asymptotic formulas (\ref{36}) and (\ref{37})
have to be applied on the paths determined by the relations $l_i \sim l_j^{1/\sigma}$,
$|V_k|\,l_k \sim  l_j^{-1/\sigma}$, $|V_m|\,l_m \sim  l_j^{-1/\sigma}$, 
$l_i \sim l_k^{1/\sigma}$, $|V_m|\,l_m \sim  l_k^{-1/\sigma}$,
$|V_j|\,l_j \sim  l_k^{-1/\sigma}$. There are no contradictions in these relations,
because, e.g., from the comparison of the relations with 
$|V_m|\,l_m$, one finds that $l_j \sim l_k$,
so that  $l_k/l_j^{1/\sigma} $ and $l_j/l_m^{1/\sigma} $ tend here to zero as required.
Thus, all the triads in the cancellation equation have to be omitted and the multiplication 
of the equation $q_it_i + q_j^2l_j + q_k^2l_k + q_m^2l_m =0$ yields the resonance equation 
\begin{equation}
\chi_{ji}^\sigma \tau_i + \eta_{jj}^\sigma +\eta_{kj}^\sigma  +
\eta_{mj}^\sigma   = 0  \,,
\label{55}
\end{equation}
which is similar to the second equation in table~\ref{tab:table3}. 
The multiplication by $l_j^{1/\sigma}$ imposes the restriction on $V_j$\,, 
namely $V_j \in {\cal G}_j^\sigma$. If some strength 
belongs to the corresponding  ${\cal G}^0$-set, say $V_m$\,, then in equation 
(\ref{55}), we have to set $\eta_{mj}^\sigma =0$. Finally, 
in a similar way, one can treat 
the situation as all the strengths are from the ${\cal G}$-sets. The multiplication
of the equation $q_it_i + q_j^2l_j + q_k^2l_k + q_m^2l_m =0$,
for instance, by $l_j^{1/\sigma}$ results in the resonance equation
\begin{equation}
\eta_{ij}^\sigma  + \eta_{jj}^\sigma +\eta_{kj}^\sigma  +\eta_{mj}^\sigma   = 0  \,,
\label{56}
\end{equation}
which is similar to the last equation in table~\ref{tab:table3}. Thus, even if all
the strengths belong to the ${\cal G}$-sets, the existence of resonance 
equations holds true on the corresponding paths with $\sigma \in [2, \infty)$.

Concerning  the double-layer structures ($N=2$), where triads (\ref{31}) 
are absent, instead of resonance equation (\ref{30}), which is valid for $\sigma =1$,
one can write the equation for the whole interval $1 < \sigma < \infty$. 
Indeed,  multiplying the equation $q_i^2l_i + q_jt_j =0$ by $l_i^{1/\sigma}$, 
 we get the resonance  equation
\begin{equation}
\lim_{(l_i,l_j) \in \Gamma^\sigma_{ij}}\!\!(q_i^2l_i + q_jt_j)\,l_i^{1/\sigma} =
\eta^\sigma_{ii} +\chi^\sigma_{ij} \tau_j =0,
\label{57}
\end{equation}
where $V_i \in {\cal G}^\sigma_{i}$ and $l_i^{1/\sigma} \sim l_j$\,. Both equations
(\ref{30}) for $\sigma =1$ and (\ref{57}) for $\sigma >1$ admit  countable sets of 
solutions if $V_j < 0$, that determine the corresponding  resonance sets.

Thus, from the above arguments, one can deduce

{\bf Conclusion 4}\,:  {\it For the existence of resonance sets, at least one of 
the strengths
 from the configuration $V_1(l_1), \ldots , V_N(l_N)$\,,  say $V_j$\,, must belong 
 to the  ${\cal G}_j$-set. In this case, for deriving a resonance equation,
 the equation for the cancellation of divergences
$\lambda_{21}^\infty =0$ can be multiplied by $l_j^{1/\sigma}$.
  During this procedure,
the asymptotic representation given by equations (\ref{36}), (\ref{37}) and (\ref{45})
has to be adopted and this application depends on  whether the middle strength $V_j$ 
in the $T_{ijk}$-triads belongs to the ${\cal G}'_j$- or ${\cal G}_j$-set. In the former
case, equations (\ref{45}) are to be used, while in the latter case, due to relations 
(\ref{36}) and (\ref{37}), the $T_{ijk}$-triads are finite in the squeezing limit 
if $\sigma \in [2, \infty )$.  In the case as one of the lateral strengths or both these
are from the ${\cal G}$-sets ($V_1 \in {\cal G}_1$ and $V_N \in {\cal G}_N$) and all the
internal strengths belong to the ${\cal G}'$-sets ($V_i \in {\cal G}'_i$, $i =2, \ldots ,
N-1$), then the resonance equations can exist on the paths with the parameter 
$\sigma$ from the interval $1 < \sigma < \infty$.
 If one of the lateral strengths or both these 
are regular, i.e., $V_1 \in {\cal G}_1^0$ and $V_N \in {\cal G}_N^0$, then  the cases
$V_1 \equiv 0$ and $V_N \equiv 0$ have to be  excluded from the consideration. 
Next, in a double-layer structure ($N=2$), 
no more than 
one strength belonging to the corresponding  ${\cal G}'$-set is available. For a single-layer 
structure ($N=1$), for the existence 
of resonance sets it is necessary that the potential must be negative with the strength
belonging  to the ${\cal G}'$-set. }

\subsection{An example}

As a particular example of the  functions $V_i(l_i) $ and $V_j(l_j)$, 
used for the realization of the singular squeezing limit, one can consider
the following asymptotic expressions: 
\begin{equation}
V_i(l_i) \sim l_i^{- \mu},~~V_j(l_j) \sim l_j^{- \nu/(1 -\mu +\nu)},~~
\mu,\, \nu >0, ~~1 -\mu +\nu >0, 
\label{58}
\end{equation}
used in several works \cite{zci,asl,zz11jpa,z18aop,zz21jpa,zz14ijmpb}.
By a proper choosing of the positive 
parameters $\mu$ and $\nu$ in (\ref{58}), all the four singular cases:   (i)
$V_i\in {\cal G}'_i $\,,  $V_j \in {\cal G}'_j $\,; (ii)
 $V_i \in {\cal G}_i \setminus {\cal G}_{i}^0$,  $V_j \in {\cal G}'_j $\,; (iii) 
 $V_i\in {\cal G}'_i $\,,   $V_j \in {\cal G}_j \setminus {\cal G}_{j}^0$\,;
  (iv) $V_i\in {\cal G}_i \setminus {\cal G}_{i}^0$,  
  $V_j \in {\cal G}_j \setminus {\cal G}_{j}^0$\,;    
 which were considered above, can be materialized on the $(\mu, \nu)$-sets
\begin{equation}
\!\!\!\!\!\!\!\!\!\!
\begin{array}{llll}
 P~  := \{\mu =\nu =2\},  & V_i \in {\cal G}'_i\,,~V_j \in {\cal G}'_j\,, \\
 L_1 := \{ 1 < \mu < 2,\, \nu = 2(\mu -1)\}, & V_i \in {\cal G}_i \setminus 
{\cal G}_i^0\,,~V_j \in {\cal G}'_j\,,  \\
 L_2 := \{\mu  =2, \, 2 < \nu < \infty\}, & V_i \in {\cal G}'_i\,,~~
~V_j \in {\cal G}_j \setminus {\cal G}_j^0\,, \\
S~ := \{1 < \mu <2,\, 2(\mu - 1)< \nu < \infty\}, 
& V_i \in {\cal G}_i \setminus {\cal G}_i^0\,,~
V_j \in {\cal G}_j \setminus {\cal G}_j^0\,.
\end{array} 
\label{59}
\end{equation}
These sets form the triangle on 
the $(\mu,\,\nu)$-plane with the vertex at the point $P$, where 
$L_1$ and $L_2$ are the edges,  and $S$ the interior of this angle. 
One can check that limits (\ref{22}), namely the first, the second, the third and the sixth 
inequalities, are fulfilled  accordingly on the paths: $\Gamma'_{ij}$  (at point $P$), 
$\Gamma'_{ij} \cup \bar\Gamma_{V_ij}$ (on  line $L_1$), 
$\Gamma'_{ij} \cup \bar\Gamma_{V_ji}$ (on  line $L_2$) and 
$\Gamma'_{ij} \cup \bar\Gamma_{V_ij} \cup \bar\Gamma_{V_ji}$ (on  plane $S$), 
where  
\begin{equation}
\begin{array}{ll}
\bar\Gamma_{V_ij} = \left\{ (l_i, l_j) \to 0~\big|~ l_i/l_j \to 0,~ l_i^{{1- \mu}}l_j
\to c \ge 0\right\}, \\
\bar\Gamma_{V_ji} = \left\{ (l_i, l_j) \to 0~\big|~ l_j/l_i \to 0,~ 
l_i\,l_j^{(1-\mu)/(1-\mu +\nu)} \to c \ge 0 \right\}.  \end{array} 
\label{60}
\end{equation}
In other words,  inequalities (\ref{22}) are fulfilled on 
 the  paths with $l_i/l_j \to c >0$ at the point $P$,
while on the lines $L_1$ and $L_2$ as well as on the plane $S$, the available paths 
are  bounded by the limit pencils
 defined by  the asymptotic relations: $l_i/l_j \to c >0$ and $l_i/l_j \sim 
l_i^{2-\mu} \to 0$ (on $L_1$); \, $l_j/l_i \sim l_j^{[2(1-\mu) +\nu]/(1-\mu +\nu)} \to 0$
 and $l_i/l_j \to c >0$ (on $L_2$);  \,
$l_j/l_i \sim l_j^{[2(1-\mu) +\nu]/(1-\mu +\nu)} \to 0$
and $l_i/l_j \sim l_i^{2-\mu} \to 0$ (on $S$).

\section{Final remarks}

Thus, we have formulated above the conditions on the strength configuration  
$V_1(l_1), \ldots , V_N(l_N)$ and the paths $l= \{l_1 , \ldots , l_N \} \to 0$, 
under which the squeezed multi-layer structure can be described 
as a one-point interaction model.
These conditions are of two types. The first of them is formulated by Conclusion 2, realizing  
 the finite and non-zero squeezed limits of 
the diagonal elements $\lambda_{11}$ and $\lambda_{22}$ of the transmission matrix $\Lambda$.

Under the conditions of the second type described in Conclusions 3 and 4, 
the squeezed $\Lambda$-matrix becomes a 
well-defined object, due to the cancellation of divergences in the singular element
$\lambda_{21}$. This cancellation imposes the constraints on the system parameters, 
resulting in the appearance of resonance sets. More precisely, 
using the equation $\det\Lambda(x_0, x_N) =1$ and $\lambda_{12} \to 0$, 
valid  for any strength configuration, one can assert that 
 $\lambda_{11} \to \theta \in \R\setminus \{0\}$ and  $\lambda_{22} \to \theta^{-1}$.
As a result, the resulting connection matrix is of the form
\begin{equation}
\Lambda_0 : = \lim_{l \to 0} \Lambda(x_0,x_N) = \left( \begin{array}{lr} \theta ~~~~0 \\
\alpha ~~~\theta^{-1} \end{array} \right),~~~~\alpha \in \R.
\label{61}
\end{equation}
As follows from equation (\ref{17}), 
the bound state  level is $\kappa = -\, \alpha/(\theta +\theta^{-1})$.

 \bigskip
{\bf  Acknowledgments}
\bigskip

The authors acknowledge a partial support by the 
National Academy of Sciences of Ukraine Grant `Functional Properties of
Materials Prospective for Nanotechnologies' (project No.~0120U100858).
They are indebted to Y.D. Golovaty for careful reading of the manuscript,
his suggestions and remarks, resulting in the improvement of the paper. 
They also thank the anonymous referees for the careful reading of this paper,
their questions and pointing out several mistakes and inaccuracies.


\bigskip
{\bf References}
\bigskip 


\begin{thebibliography}{99}


\bibitem{a-h}
Albeverio~S, Gesztesy~F,  H{\o}egh-Krohn~R and Holden~H 2005 
 {\it Solvable Models in Quantum Mechanics (With an Appendix by Pavel Exner)} 
2nd revised edn  (Providence: RI: American Mathematical Society: Chelsea Publishing)

\bibitem{ak}
Albeverio~S and Kurasov~P 1999 {\it Singular Perturbations of Differential Operators: 
Solvable Schr\"{o}dinger-Type Operators} (Cambridge: Cambridge University Press) 


\bibitem{s}
\v{S}eba~P 1986 {\it Rep. Math. Phys.} {\bf 24} 111

\bibitem{gm}
 Golovaty~Y~D and Man'ko~S~S 2009 {\it Ukr. Math. Bull.}
 {\bf 6} 169  

\bibitem{gh}
Golovaty~Y~D and Hryniv~R~O 2010 {\it J. Phys. A: Math. Theor.}
 {\bf 43} 155204 \\
Golovaty~Y~D and Hryniv~R~O 2010 {\it J. Phys. A: Math. Theor.}
 2011 {\bf 44} 049802  (corrigendum)

\bibitem{gh1}
Golovaty~Y~D and Hryniv~R~O 2013 {\it Proc. R. Soc. Edinb.~A}
 {\bf 143} 791   


\bibitem{m1}
Man'ko~S~S 2010 {\it J. Phys. A: Math. Theor.} {\bf 43} 445304

\bibitem{m2}
Man'ko~S~S 2012  {\it J. Math. Phys.} {\bf 53} 123521

\bibitem{em1}
Exner~P and Manko~S~S 2013 {\it J. Phys. A: Math. Theor.} {\bf 46} 345202 


\bibitem{tn}
 Toyama~F~M and  Nogami~Y 2007 {\it J. Phys. A: Math. Theor.} {\bf 40} F685   


\bibitem{c-g}
 Christiansen~P~L, Arnbak~N~C, Zolotaryuk~A~V, Ermakov~V~N and 
Gaididei~Y~B 2003 {\it J. Phys. A: Math. Gen.} {\bf 36} 7589 

\bibitem{zci}
 Zolotaryuk~A~V, Christiansen~P~L and Iermakova~S~V 2006 
{\it J. Phys. A: Math. Gen.} {\bf 39} 9329

\bibitem{asl}
Zolotaryuk~A~V 2008 {\it J. Comput. Theor. Nanoscience } {\bf 1} 187


\bibitem{zz11jpa}
Zolotaryuk~A~V and Zolotaryuk~Y 2011 {\it J. Phys. A: Math. Theor.} 
{\bf 44} 375305 \\
Zolotaryuk~A~V and Zolotaryuk~Y 2012 {\it J. Phys. A: Math. Theor.} 
{\bf 45} 119501 (corrigendum)

\bibitem{g1}
Golovaty~Y 2012 {\it Methods Funct. Anal. Topology} {\bf 18} 243

\bibitem{g2}
Golovaty~Y 2013 {\it Integr. Equ. Oper. Theor.} {\bf 75} 341


\bibitem{gnn}
Gadella~M, Negro~J and  Nieto~L~M 2009 {\it Phys. Lett.~A}  {\bf 373} 1310 


 \bibitem{gmmn}
 Gadella~M,  Mateos-Guilarte~J, Mu\~{n}oz-Casta\~{n}eda~J~M and Nieto~L~M
 2016  {\it J. Phys. A: Math. Theor.}  {\bf 49} 015204


\bibitem{ggn}
 Gadella~M,  Glasser~M~L and Nieto~L~M 2011 
{\it Int. J. Theor. Phys.}  {\bf 50} 2144

\bibitem{gggm}
 Gadella~M,   Garc\'{i}a-Ferrero~M~A, 
 Gonz\'{a}lez-Mart\'{i}n~S and  Maldonado-Villamizar~F~H 2014
 {\it Int. J. Theor. Phys.} {\bf 53} 1614   

\bibitem{fggn1}
 Fassari~S,  Gadella~M,  Glasser~M~L and  Nieto~L~M  2018 
{\it Ann. Phys. (NY)} {\bf 389} 48


\bibitem{adk}
Albeverio~S, D\c{a}browski~L and Kurasov~P 1998 {\it Lett. Math. Phys.} {\bf 45} 33

\bibitem{an1}
Albeverio~S and Nizhnik~L 2007 {\it J. Math. Anal. Appl.} {\bf 332} 884

\bibitem{l1}
 Lange~R-J 2012 {\it J. High Energy Phys.} {\bf JHEP11(2012)}032  
 
 \bibitem{bn}
Brasche~J~F and Nizhnik~L~P 2013 {\it Methods Funct. Anal. Topology} {\bf 19} 4

\bibitem{an2}
Albeverio~S and Nizhnik~L 2013 {\it Methods Funct. Anal. Topology} {\bf 19} 199

   \bibitem{l2}
 Lange~R-J 2015  {\it J. Math. Phys.} {\bf 56} 122105

\bibitem{g3}
Golovaty~Y  2018 {\it J. Phys. A: Math. Theor.}  {\bf 51} 255202  


\bibitem{cs}    
  Cheon~T and Shigehara~T 1998 {\it Phys. Lett.~A} {\bf 243} 111

\bibitem{enz}
Exner~P, Neidhardt~H and Zagrebnov~V~A 2001
 {\it Commun. Math. Phys.} {\bf 224}  593


\bibitem{afr1}
 Albeverio~S, Fassari~S and Rinaldi~F 2013 {\it J. Phys. A: Math. Theor.} 
  {\bf 46} 385305
 
 \bibitem{afr2}
 Albeverio~S, Fassari~S and  Rinaldi~F 2016 {\it J. Phys. A: Math. Theor.} 
  {\bf 49} 025302

 
\bibitem{z18pe} 
Zolotaryuk~A~V 2018 {\it Physica E: Low-dimensional Systems and Nanostructures}
 {\bf 103} 81 
 
  \bibitem{z18aop}
 Zolotaryuk~A~V 2018 {\it Ann. Phys. (NY)} {\bf 396} 479


\bibitem{zz20ltp}
Zolotaryuk~A~V and Zolotaryuk~Y 2020 {\it Low Temperature Phys.} 2020 {\bf 46} 927

\bibitem{zz21jpa}
Zolotaryuk~A~V and Zolotaryuk~Y 2021 {\it J. Phys. A: Math. Theor.} 
{\bf 54}  035201


\bibitem{k}
Kurasov~P 1996 {\it J. Math. Anal. Appl.} {\bf 201} 297

\bibitem{cnf}
Coutinho~F~A~B, Nogami~Y and P\'{e}rez~J~F 1997 {\it J. Phys. A: Math. Gen.} {\bf 30} 3937


\bibitem{zz14ijmpb} 
Zolotaryuk~A~V and Zolotaryuk~Y 2014 {\it Int. J. Mod. Phys. B} {\bf 28} 1350203
 



\end{thebibliography}
 \end{document}